\documentclass[aps,pra,reprint,superscriptaddress,amsmath,amssymb,floatfix,nofootinbib]{revtex4-2}
\usepackage[T1]{fontenc}
\usepackage[utf8]{inputenc}
\usepackage{braket,graphicx,xcolor}
\usepackage[colorlinks=true,allcolors=blue]{hyperref}
\hypersetup{pdftitle={Homodyne versus heterodyne detection near Gaussian states: Fraction of the quantum Fisher information recovered},pdfauthor={Franciszek Grzywa, Sebastian Dolata, Karol Bartkiewicz}}

\newcommand{\nhat}{\hat n}
\newcommand{\Sym}{\mathrm{Sym}}
\newcommand{\wick}[1]{{:}#1{:}}
\DeclareMathOperator{\tr}{Tr}
\newtheorem{theorem}{Theorem}
\let\citep\cite
\begin{document}
\title{Homodyne versus heterodyne detection near Gaussian states:\texorpdfstring{\\}{ } Fraction of the quantum Fisher information recovered}
\author{Franciszek Grzywa}
\author{Sebastian Dolata}
\author{Karol Bartkiewicz}\email[Contact author: ]{bark@amu.edu.pl}
\affiliation{Institute of Spintronics and Quantum Information, Faculty of Physics and Astronomy, Adam Mickiewicz University, 61-614 Pozna\'n, Poland}
\date{\today}
\begin{abstract}
Homodyne and heterodyne detection are the standard measurements of an optical mode. Which of them yields more information is known only for particular tasks, with opposite answers for displacements and tomography.
We compare them for every small change of a Gaussian state by their efficiency, the recovered fraction of the quantum Fisher information.
We prove that near any Gaussian state of any number of modes the quantum Fisher information and the Fisher information of any Gaussian measurement are sums over the Hermite orders of the change, each term fixed by the covariance matrix and the measurement.
A fixed Gaussian measurement measures no change more efficiently than the best displacement.
Near a thermal state of one mode closed-form efficiencies reconcile the known comparisons, and homodyne detection at a uniformly random phase recovers at most one half, a bound raised by squeezing.
Near a coherent state both detections recover exactly one half for every change of finite quantum Fisher information (the largest possible average).
At the two-mode squeezed vacuum, homodyne detection of both modes at uniformly random phases is as efficient as heterodyne detection for every such change.
On simulated records of one set of distant states, a likelihood-ratio observer shows a heterodyne advantage under Gaussian noise and none under loss, as the theory predicts. Under dephasing it shows none.
Neural networks on reconstructed Wigner functions can show a heterodyne advantage absent from the records.
For random non-Gaussian changes of up to seven photons near a thermal state the median efficiencies of both detections lie between 0.04 and 0.19, 5 to 25 times below the optimal measurement.
\end{abstract}
\maketitle
\section{Introduction}\label{sec:intro}

\begin{figure*}[!t]
\centering\includegraphics[width=\textwidth]{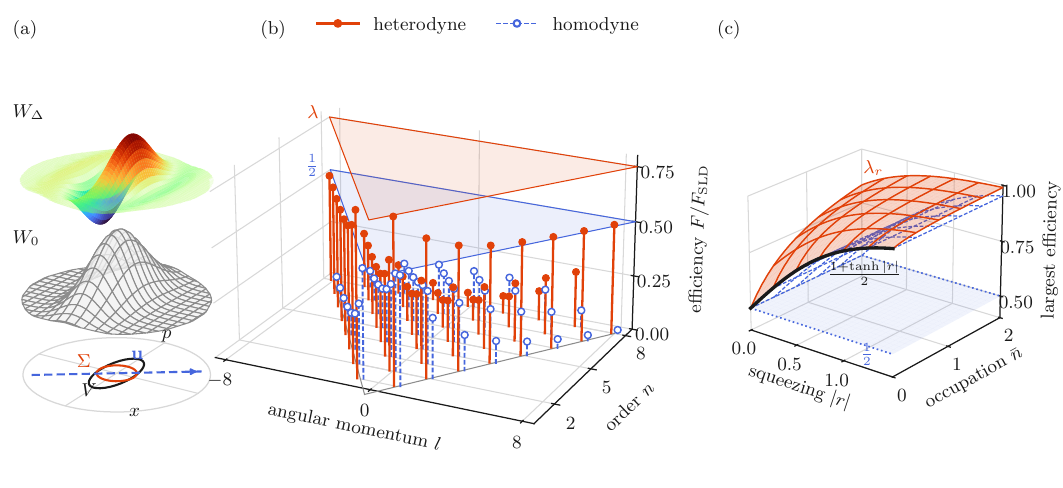}
\caption{Information that Gaussian measurements recover near a Gaussian state. (a) Objects of the theory in the phase space of the quadratures $x$ and $p$. Middle: Wigner function $W_0$ of the reference state, here a squeezed thermal state. Top: a change $W_\Delta=g_nW_0$ of the state in one Hermite--Gauss mode $g_n$ (Sec.~\ref{sec:fisher-modes}), here of order $n=3$ and angular momentum $l=1$. Bottom: the covariance ellipse $V$ of the state, the noise $\Sigma$ that heterodyne detection adds (circle), and the quadrature $\mathbf u$ measured by homodyne detection (dashed arrow). (b) Efficiency $F/F_{\mathrm{SLD}}$ for every mode $(n,l)$ with $n\le8$ near a thermal state with occupation $\bar n=1$ (Sec.~\ref{sec:fisher-thermal}). The efficiency is the classical Fisher information of the detection divided by the quantum Fisher information. Filled markers, solid stems: heterodyne detection. Open markers, dashed stems: homodyne detection at a uniformly random phase. The shaded planes show the bounds for all directions, $\lambda=(2\bar n+1)/(2\bar n+2)$ and $1/2$, which are attained by the displacements ($n=1$). (c) Largest efficiency over all directions near a squeezed thermal state, plotted against the squeezing parameter $|r|$ and the occupation $\bar n$ (Sec.~\ref{sec:fisher-squeezed}). Solid mesh: heterodyne detection, $\lambda_r$ of Eq.~\eqref{eq:sqhet}. Dashed mesh: homodyne detection at a uniformly random phase, $(1+\tanh|r|)/2$. For a squeezed vacuum the two bounds coincide (thick line). Shaded plane: homodyne detection at a random phase that is uniform in the coordinates in which the reference state is isotropic, $1/2$.}\label{fig:overview}
\end{figure*}

Homodyne and heterodyne detection are the two standard measurements of an optical mode. They are used in experiments to characterize Gaussian states and to check sources of non-Gaussian states, such as Fock, cat and Gottesman--Kitaev--Preskill (GKP) states~\citep{GKP2001,Ourjoumtsev2007,Konno2024,Larsen2025}. Such states are resources for quantum communication and fault-tolerant computation~\citep{Terhal2020}. Homodyne detection records one quadrature at a chosen phase. In homodyne tomography the records are taken at several phases and inverted into a Wigner function or a density matrix~\citep{LvovskyRaymer2009,VogelRisken1989,Smithey1993}. Heterodyne detection, on the other hand, records both quadratures at once, with one additional unit of vacuum noise, and the outcomes are samples of the Husimi $Q$ function~\citep{ArthursKelly1965,LeonhardtPaul1995}. We note that both detections are Gaussian measurements and that, for a slight change of the state, the information recovered by any measurement is limited by the quantum Fisher information. In this paper we want to know how close the two standard detections come to this limit and for which changes of the state. We consider an arbitrary weak signal imprinted on a Gaussian probe, whether this signal is Gaussian or not. For such a signal our answer also gives the number of preparations that the two detections need in comparison with the measurement which is optimal for it.

So far, which of the two measurements carries more information about the state has been established only for particular tasks. With single-mode probes heterodyne detection is optimal for the joint estimation of both components of a displacement~\citep{YuenLax1973,Holevo1982,Genoni2013}. It was also found to estimate the first and second quadrature moments more accurately for almost all Gaussian states and for several non-Gaussian states~\citep{Rehacek2015,Muller2016,Teo2017}. The quantum Fisher information of Gaussian probes has a closed form~\citep{Monras2013,Pinel2013,Safranek2019}. For weak perturbations of a Gaussian state that leave its covariance matrix unchanged, it is bounded from above by the non-Gaussianity of the perturbed state~\citep{Genoni2011geo}. In phase estimation homodyne detection reaches the quantum Fisher information for displaced thermal and squeezed vacuum probes, but not for every Gaussian probe~\citep{Aspachs2009,Oh2019}. In thermometry Gaussian measurements approach the quantum limit only at high temperature~\citep{Cenni2022}. For full tomography the authors of Ref.~\cite{Fernandes2025} found that homodyne detection gives a lower Cram\'er--Rao bound on the reconstruction error for every state they tested, Gaussian states included. The worst-case upper bounds on the number of copies grow as a high power of the photon number for homodyne detection, whereas for heterodyne detection their growth is effectively exponential~\citep{Gandhari2024}. For two-mode squeezed probes the Fisher information of double homodyne detection has been computed for a displacement of one mode and for the squeezing parameter~\citep{Bressanini2024}. For loss in one arm of such probes the Fisher information is known for several Gaussian measurements~\citep{Radia2026}. Each of these comparisons refers to the parameter or the figure of merit of one task only.

The quantum Fisher information bounds the classical Fisher information of every measurement on single copies~\citep{Helstrom1976,BraunsteinCaves1994,Paris2009}, and we take it as a natural reference for this comparison. In finite dimensions there exist measurements that recover one half of it in every direction, i.e., for every small change of a pure state. Such measurements are called Fisher-symmetric, and they are optimal in the sense of the Gill--Massar bound~\citep{GillMassar2000,LiFerrie2016,ZhuHayashi2018}. For Gaussian measurements the recovered fraction has been obtained in special cases, namely for displacements~\citep{YuenLax1973,Holevo1982,Genoni2013} and for the temperature of a thermal state~\citep{Cenni2022}. It has also been shown that for thermal states under unitary signals the Fisher information of the Husimi function bounds the quantum Fisher information from below~\citep{Chitgopekar2026}. Near a thermal state the Hermite--Gauss modes that we use below have been studied as well. They form the damping basis of an oscillator damped by a thermal bath~\citep{BriegelEnglert1993,EnglertMorigi2002} and coincide with the eigenvectors of the quantum Ornstein--Uhlenbeck semigroup~\citep{Cipriani2000}, which obeys detailed balance~\citep{Alicki1976,CarlenMaas2017}. The norms of the eigenvectors in the metric of the quantum Fisher information are not given in these works. On the classical side of the problem it is known that the conditional expectation of jointly Gaussian variables acts on each Hermite order as a symmetric power~\citep{Janson1997}. From this property one obtains the Hermite singular functions of the Gaussian channel~\citep{AbbeZheng2012} and the singular value decomposition of the Radon transform with a Gaussian weight~\citep{Davison1981,Louis1985}. Nevertheless, to our knowledge, the fractions of the quantum Fisher information that Gaussian measurements recover along the Hermite--Gauss modes have not been reported beyond these special cases. The same holds for bounds on these fractions over all directions near a Gaussian state, in particular for directions that change the non-Gaussian structure of the state.

In this paper we determine how much information Gaussian measurements recover near a Gaussian state. Our general result holds for any Gaussian state of any number of modes. We evaluate the recovered fraction in closed form for thermal and coherent states of one mode and bound it for squeezed thermal states (Fig.~\ref{fig:overview}). As an example of several modes we treat the two-mode squeezed vacuum. To this end we expand the change of the Wigner function, divided by the Gaussian Wigner function of the reference state, in Hermite--Gauss modes. These modes are functions on phase space (they are not transverse modes of a beam). The order $n$ of a mode measures how fine its structure in phase space is, while its angular momentum $l$ tells how fast the mode oscillates with the phase. Our main result is the efficiency of a Gaussian measurement, i.e., the fraction of the quantum Fisher information that it recovers, for every change and in every order. It rests on a property that we prove in Theorem~\ref{thm:structure}: neither the quantum Fisher information nor the Fisher information of a Gaussian measurement couples modes of different order. In the order $n$ both quantities are given by the theorem in terms of the $n$th symmetric powers of matrices of the size of phase space ($2\times2$ for one mode). The classical half of this statement follows from known properties of Gaussian variables~\citep{Janson1997}. As regards the quantum half, it is the mean of two members of a recently reported family of inner products~\citep{Li2026} (Sec.~\ref{sec:fisher-main}). The new element of the theorem is their comparison. From it we obtain the efficiency in every order from an eigenvalue problem, whose dimension is $n+1$ for one mode.

From this structure we derive the following results. (i) We show that a fixed Gaussian setting of any number of modes measures no change of the state more efficiently than the best displacement (Theorem~\ref{thm:single}). In Theorem~\ref{thm:mix} we consider settings of one mode drawn at random, such as homodyne detection with any distribution of the phase. For such settings we bound the efficiency by the Fourier coefficients of that distribution. (ii) Near a thermal state of one mode with the occupation (mean photon number) $\bar n$ every efficiency has a closed form (Sec.~\ref{sec:fisher-thermal}). Homodyne detection at a uniformly random phase that is recorded with each event, as in homodyne tomography, recovers at most one half of the quantum Fisher information. Heterodyne detection recovers at most $(2\bar n+1)/(2\bar n+2)$, and both bounds are attained by displacements. (iii) We derive a criterion on the order and the angular momentum that tells mode by mode which detection carries more information (Sec.~\ref{sec:fisher-thermal}). For edge modes ($|l|=n$), and at larger occupation also for low orders, heterodyne detection carries more. Homodyne detection, in turn, carries more for central modes ($|l|\le1$) of high order, i.e., for modes which describe fine radial and interference structure. With this criterion we reconcile the known comparisons for particular tasks. (iv) Near a coherent state both detections recover exactly one half in every direction of finite quantum Fisher information. They are thus Fisher-symmetric measurements at coherent states (Sec.~\ref{sec:fisher-squeezed}). (v) Near a squeezed thermal state both bounds grow with the squeezing parameter $r$. Homodyne detection at a uniformly random phase then recovers at most the fraction $(1+\tanh|r|)/2$ [Fig.~\ref{fig:overview}(c)]. (vi) Since the Fisher informations are $\chi^2$ divergences of the outcome distributions, the bounds hold at any distance from a mixed reference state (Sec.~\ref{sec:fisher-chi2}). Near a thermal state the comparison of the two detections in each mode is changed by loss and Gaussian noise only through the occupation (Sec.~\ref{sec:fisher-noise}).

(vii) At the two-mode squeezed vacuum, we show that heterodyne detection of both modes and homodyne detection of both modes at uniformly random phases recover the same information in every direction of finite quantum Fisher information. Their efficiencies in the order $n$ are $(1\pm\tanh^nr)/2$. For a measurement of one arm half of the efficiencies of every order vanish, and the other half are those of the thermal state of this arm. We also find that no nonadaptive Gaussian measurement of each mode separately is Fisher-symmetric (Secs.~\ref{sec:fisher-pure} and~\ref{sec:fisher-tmsv}).

(viii) We test how far the theory of small changes carries for distant states on the identification of seven families of non-Gaussian and Gaussian states from finite records of events (Sec.~\ref{sec:test}). There we compare the two detections by quantities that involve no trained classifier, i.e., by the Chernoff exponents of the outcome distributions and by a likelihood-ratio observer, which makes the optimal decision on a record. The Chernoff exponents of the two detections differ little for the median pair of states. In the base set the observer on records of $100$ events shows no heterodyne advantage under loss and dephasing, whereas under Gaussian noise it shows one. The theory predicts the effect of dephasing on a pair of states as long as dephasing leaves the mean outcome distribution of the pair almost unchanged. In contrast, neural networks that classify a reconstructed Wigner function show a heterodyne advantage in the base set under loss and dephasing, where the observer shows none. This advantage disappears for a network trained on the homodyne records themselves. We use the networks only to demonstrate how a classifier used in practice relates to the theory, and the comparison of the two detections is not based on them.

For experiments, our results give the price of the two standard detections in the number of preparations, relative to the measurement that is optimal for a known change. Near thermal and coherent states this price is a factor of at least two for homodyne detection at a uniformly random phase. For random non-Gaussian changes of up to seven photons near a thermal state it is about $5$ to $25$. Once the change of interest and the noise of the state are given, the results also tell which of the two detections carries more information.

Section~\ref{sec:fisher} contains the theory. We present the reference states and the measurements in Sec.~\ref{sec:fisher-setting}, the Hermite--Gauss modes in Sec.~\ref{sec:fisher-modes}, the main result in Sec.~\ref{sec:fisher-main} and the bounds in Sec.~\ref{sec:fisher-bounds}. We then treat thermal states in Sec.~\ref{sec:fisher-thermal}, coherent and squeezed states in Sec.~\ref{sec:fisher-squeezed}, pure states of several modes in Sec.~\ref{sec:fisher-pure} and the two-mode squeezed vacuum in Sec.~\ref{sec:fisher-tmsv}. Distant states follow in Sec.~\ref{sec:fisher-chi2} and noise in Sec.~\ref{sec:fisher-noise}. Section~\ref{sec:test} tests the theory on an identification task, Sec.~\ref{sec:discussion} discusses the results, and Sec.~\ref{sec:conclusions} concludes. Appendix~\ref{app:proofs} contains the conventions, the proofs of the three theorems, the numerical checks and the derivations for distant states and noise channels.
\section{Fisher information of Gaussian measurements}\label{sec:fisher}

We quantify here the information about a change of the state contained in a single detection event, independently of any classifier. Let a source prepare a state $\rho_\varepsilon=\rho_0+\varepsilon\Delta+O(\varepsilon^2)$ instead of a reference state $\rho_0$, where $\Delta$ is Hermitian and traceless. For a pure $\rho_0$ the operator $\rho_0+\varepsilon\Delta$ need not be a state, and the term of the second order keeps $\rho_\varepsilon$ a state. For the measurements considered here this term does not change the relations below. Their outcome densities for $\rho_0$ vanish nowhere (Appendix~\ref{app:local}). We call $\Delta$ the direction of the change. Suppose that we collect $N$ independent detection events of a fixed measurement, each from one preparation (copy) of the state. Then the error of the optimal decision between $\rho_0$ and $\rho_\varepsilon$ falls as $e^{-N\xi}$~\citep{Chernoff1952}. The exponent is the Chernoff information
\begin{equation}
\xi=-\ln\min_{0\le s\le1}\int f_0^{\,s}f_1^{\,1-s}
\label{eq:chernoff}
\end{equation}
of the outcome densities $f_0$ and $f_1$ of a single event for the two states. We also call $\xi$ the Chernoff exponent. If $f=f_0$ is the outcome density for $\rho_0$ and $\delta$ its change for the direction $\Delta$, then
\begin{equation}
\xi=\frac{\varepsilon^2}{8}F+O(\varepsilon^3),\qquad F(\Delta)=\int\frac{\delta^2}{f},
\label{eq:localchernoff}
\end{equation}
where $F$ is the classical Fisher information of the measurement in the direction $\Delta$ (Appendix~\ref{app:local}). The quadratic form $F$ also has a meaning for a large change. For any state $\rho$, $F(\rho-\rho_0)$ is the $\chi^2$ divergence of the outcome distribution of $\rho$ from that of $\rho_0$ (Sec.~\ref{sec:fisher-chi2}).

\subsection{Gaussian states and Gaussian measurements}\label{sec:fisher-setting}
As reference states we take the Gaussian states of $M$ modes. Let $\hat x_j=(\hat a_j+\hat a_j^\dagger)/\sqrt2$ and $\hat p_j=-i(\hat a_j-\hat a_j^\dagger)/\sqrt2$ be the quadratures of the mode $j$ with the annihilation operator $\hat a_j$ (with $\hbar=1$, so that the vacuum variance is $1/2$), and let $\mathbf x=(x_1,p_1,\dots,x_M,p_M)$ be a point of phase space. The Wigner function $W_0$ of a Gaussian state $\rho_0$ is a normal density with the $2M\times2M$ covariance matrix $V$. This matrix obeys $V+i\Omega/2\ge0$, where the symplectic form $\Omega$ is the direct sum of $M$ blocks $\big(\begin{smallmatrix}0&1\\-1&0\end{smallmatrix}\big)$. Since all measurements considered below are covariant under displacements, the mean of $W_0$ does not change any of the results and we set it to zero. The state $\rho_0$ is brought by a Gaussian unitary operation to a product of thermal states of $M$ normal modes with the occupations $\bar n_j$. The invariants of $V$ under such operations are the symplectic eigenvalues $v_j=\bar n_j+1/2$ (Appendix~\ref{app:conv}). We call a normal mode pure if $v_j=1/2$. For one mode we omit the index $j$, and then $v=\sqrt{\det V}=\bar n+1/2$. A thermal state has $V=v\,\openone$, where $\openone$ is the identity matrix of the dimension given by the context. For a coherent state $v=1/2$, and a squeezed thermal state with the squeezing parameter $r$ has the quadrature variances $ve^{\mp2r}$. We also use $q=\bar n/(\bar n+1)$, $\lambda=(2\bar n+1)/(2\bar n+2)$, $\kappa=2\bar n/(2\bar n+1)$ and $\tau=1/(2v)=1-\kappa$.

We consider the following measurements. The outcome of a Gaussian measurement is a vector $\mathbf y$ of $d$ real numbers that is distributed as $H\mathbf X+\mathbf N$. Here $\mathbf X$ is distributed with the Wigner function $W$ of the state, and the real $d\times2M$ matrix $H$ selects the measured combinations of the quadratures. The noise added by the measurement, $\mathbf N$, is an independent normal vector with zero mean and the covariance matrix $\Sigma$. We require
\begin{equation}
\Sigma\pm\tfrac i2H\Omega H^{\mathsf T}\ge0,
\label{eq:phys}
\end{equation}
where ${}^{\mathsf T}$ denotes the transpose. This condition guarantees that the distribution belongs to a measurement. It cannot be dropped, as we show in Appendix~\ref{app:bounds}. In a general-dyne measurement~\citep{GenoniLamiSerafini2016} the modes are projected onto displaced copies of a pure Gaussian state, which we call the seed. We use the same name for a mixed seed, for which the measurement operators are displaced copies of the seed (the noisy general-dyne measurement of Ref.~\cite{GenoniLamiSerafini2016}). In this case $H=\openone$ and $\Sigma$ is the covariance matrix of the seed ($\Sigma+i\Omega/2\ge0$). The outcome is then a phase-space point, distributed as $W$ convolved with a normal density of covariance $\Sigma$. Heterodyne detection of every mode is the case $\Sigma=\openone/2$, for which the outcome density is the Husimi function $Q$. Homodyne detection measures commuting quadratures without added noise ($\Sigma=0$ and $H\Omega H^{\mathsf T}=0$). For one mode it measures the quadrature $\mathbf u\cdot\hat{\mathbf x}$, with $\hat{\mathbf x}=(\hat x,\hat p)$, a unit vector $\mathbf u=(\cos\theta,\sin\theta)$ and $H=\mathbf u^{\mathsf T}$. It is the limit of infinite squeezing of $\Sigma$ along $\mathbf u$, and its outcome density is the marginal of $W$ along $\mathbf u$ [Fig.~\ref{fig:overview}(a)]. For several modes the rows of $H$ can select one quadrature of every mode (local homodyne detection), joint quadratures of several modes, or the quadratures of a subset of the modes. Finally, the setting ($H$ and $\Sigma$, e.g., the phase $\theta$) can be drawn at random and recorded with each event, as in homodyne tomography. The Fisher information is then the average over the settings. Since the Fisher information of independent events is additive, the same average applies if the settings follow a fixed schedule, e.g., a phase that is scanned evenly over $[0,\pi)$. We write $F_Q$ for heterodyne detection of every mode and $F_W$ for homodyne detection of every mode at phases drawn independently and uniformly from $[0,\pi)$. The subscripts name the phase-space function to which each detection gives access. Nevertheless, $F_W$ is the information of the raw quadrature values, and no reconstruction of $W$ is involved.

No measurement on single copies has a larger Fisher information than the quantum Fisher information of the symmetric logarithmic derivative (SLD)~\citep{Helstrom1976,BraunsteinCaves1994,Paris2009},
\begin{equation}
F(\Delta)\le F_{\mathrm{SLD}}(\Delta)=2\sum_{m,m'}\frac{|\Delta_{mm'}|^2}{\pi_m+\pi_{m'}},
\label{eq:fsld}
\end{equation}
where $\pi_m$ are the eigenvalues of $\rho_0$ and $\Delta_{mm'}$ the elements of $\Delta$ in its eigenbasis. This is $\tr(\Delta L)$ for the SLD $L$ defined by $\Delta=(\rho_0L+L\rho_0)/2$. We call the ratio $F/F_{\mathrm{SLD}}$ the efficiency of a measurement in the direction $\Delta$. Note that it is not the quantum efficiency of the photodiodes, which we include in the loss channel of Sec.~\ref{sec:fisher-noise}.

\subsection{Hermite--Gauss modes}\label{sec:fisher-modes}
\begin{figure}[t]
\centering\includegraphics[width=\columnwidth]{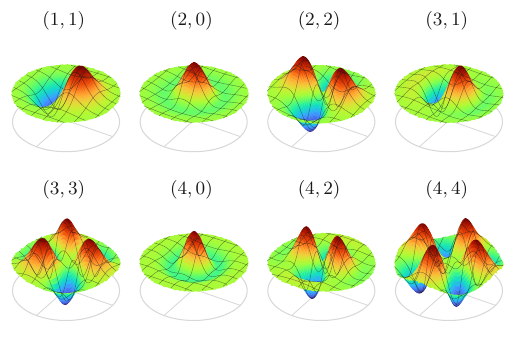}
\caption{Hermite--Gauss modes of one mode near a thermal state. Each panel shows the change $W_\Delta=gW_0$ of the Wigner function for the real part of the mode $(n,l)$ of order $n$ and angular momentum $l$, as a surface over the plane of the normalized quadratures, within the radius $3.6$ around the origin. Its value, in units of its largest modulus, is given by the height and by the color (red: positive, green: zero, blue: negative). The mode $(1,1)$ is a displacement and $(2,2)$ a squeezing, while $(2,0)$ is a change of the temperature. The modes with $|l|=n$ are the edge modes and those with $|l|\le1$ the central modes.}\label{fig:modes}
\end{figure}

We describe a direction $\Delta$ by the change $W_\Delta$ of the Wigner function divided by that of the reference state, $g=W_\Delta/W_0$, and we use the inner product $\langle g,h\rangle=\int \bar g\,h\,W_0\,d^{2M}x$. The functions of $\mathbf x$ that are square integrable in this sense decompose into Hermite orders. The order $n$ contains the polynomials of degree $n$ that are orthogonal to all polynomials of lower degree. We write $g=\sum_{n\ge1}g_n$ and call the components the Hermite--Gauss modes of order $n$ of the direction (the order $0$ is absent because $\Delta$ is traceless). The order $n$ measures how fine the structure of a mode is. In the first order we have the linear functions, which describe displacements. The changes of the covariance matrix belong to the second order, while the non-Gaussian changes are described by the higher orders. For example, near a thermal state of one mode the first order is spanned by $x$ and $p$, and the second order by $x^2-p^2$, $xp$ and $x^2+p^2-2v$. The first two functions of the second order describe a squeezing of the state, whereas the third one corresponds to a change of its temperature (the modes $(2,\pm2)$ and $(2,0)$ of Fig.~\ref{fig:modes}). Only constants such as $2v$ are fixed by the orthogonality to the lower orders, in the same way as for the Hermite polynomials of one variable. The modes reduce to these polynomials if $g$ depends on one quadrature only.

The orders are related to each other in the same way as the sectors of $n=1,2,\dots$ photons of a set of bosonic modes, and we use this analogy throughout. As the basis of the first order we take the $2M$ normalized quadratures $\mathbf z=V^{-1/2}\mathbf x$, i.e., the quadratures in units of the noise of the reference state. For a thermal state of one mode $\mathbf z=\mathbf x/\sqrt v$, and for a squeezed state the squeezed and the antisqueezed quadrature are each divided by their own standard deviation. In these coordinates the Wigner function of the reference state is isotropic with unit variance, and the functions $\mathbf c\cdot\mathbf z$ with unit vectors $\mathbf c$ are orthonormal. All matrices below are written in these coordinates. The modes of order $n$ are built from the products of $n$ linear functions, just as the states of $n$ photons in $2M$ optical modes are built from the products of $n$ creation operators. Since in both cases the order of the factors does not matter, the space of order $n$ has the dimension $\binom{n+2M-1}{n}$. This is the number of ways to distribute $n$ photons over $2M$ modes, and for one mode it equals $n+1$~\citep{Janson1997}.

If the linear functions are transformed by a $2M\times2M$ matrix $B$, every product of $n$ of them is transformed by $B$ in each of its factors. We denote the matrix that results in the order $n$ by $\mathrm{Sym}^n(B)$ and call it the $n$th symmetric power of $B$. For a unitary $B$ it is the matrix by which a linear optical network with the mode transformation $B$ acts on the states of $n$ photons. Of its properties we need only $\mathrm{Sym}^1(B)=B$, the product rule $\mathrm{Sym}^n(BC)=\mathrm{Sym}^n(B)\,\mathrm{Sym}^n(C)$, and the relation between the eigenvalues. If $B$ has the eigenvalues $\lambda_i(B)$, the eigenvalues of $\mathrm{Sym}^n(B)$ are all products of $n$ of them, $\lambda_{i_1}(B)\lambda_{i_2}(B)\cdots\lambda_{i_n}(B)$ with $i_1\le i_2\le\dots\le i_n$.

For one mode a convenient basis of the order $n$ consists of the $n+1$ modes with the angular momentum $l\in\{-n,-n+2,\dots,n\}$, which in the coordinates $\mathbf z$ oscillate as $e^{-il\vartheta}$ with the polar angle $\vartheta$ and have $k_0=(n-|l|)/2$ radial nodes (Appendix~\ref{app:conv}). They are proportional to $|\mathbf z|^{|l|}L_{k_0}^{(|l|)}(|\mathbf z|^2/2)\,e^{-il\vartheta}$, where $L_{k_0}^{(|l|)}$ is an associated Laguerre polynomial, i.e., they have the form of the Laguerre--Gauss modes of a paraxial beam. In the analogy with photons, the two functions $z_1\mp iz_2$ play the role of two bosonic modes, and the mode of angular momentum $l$ contains $(n+l)/2$ factors of the first function and $(n-l)/2$ factors of the second. We call the modes with $|l|=n$ the edge modes and those with $|l|\le1$ the central modes. Figure~\ref{fig:modes} shows the modes of the orders $n\le4$.

\subsection{Main result}\label{sec:fisher-main}
In our main result we express the classical and the quantum Fisher information near any Gaussian state, in every order, by two $2M\times2M$ matrices. In the normalized quadratures they read
\begin{equation}
\begin{split}
\Lambda&=V^{1/2}H^{\mathsf T}\big(HVH^{\mathsf T}+\Sigma\big)^{-1}HV^{1/2},\\
G_\pm&=V^{-1/2}\Big(V\pm\frac i2\Omega\Big)V^{-1/2}.
\end{split}
\label{eq:LamG}
\end{equation}
With the matrix $\Lambda$ we compare the noise of the state with the noise added by the measurement. Its eigenvalues lie between $0$ and $1$, and each of them gives the fraction of the variance of the outcomes that comes from the state. An eigenvalue equals $1$ for a quadrature that is measured without added noise. For a quadrature that is not measured it is $0$. For general-dyne detection this matrix is $V^{1/2}(V+\Sigma)^{-1}V^{1/2}$. In the case of homodyne detection it is the projector onto the measured quadratures, which for one mode reads $\Lambda=V^{1/2}\mathbf u\mathbf u^{\mathsf T}V^{1/2}/(\mathbf u^{\mathsf T}V\mathbf u)$. The matrices $G_\pm$ are those of the uncertainty relation $V\pm i\Omega/2\ge0$, and their eigenvalues are $1\pm1/(2v_j)$. An eigenvalue vanishes if and only if a normal mode is pure, i.e., if the state saturates the uncertainty relation in this mode.

\begin{theorem}[structure]\label{thm:structure}
Let $\rho_0$ be a Gaussian state of $M$ modes with the covariance matrix $V$ and let $\Delta$ be a direction with finite $F_{\mathrm{SLD}}(\Delta)$ and Hermite--Gauss modes $g_n$. Then the Fisher information of a Gaussian measurement that obeys Eq.~\eqref{eq:phys} and the quantum Fisher information do not couple different orders,
\begin{equation}
F(\Delta)=\sum_{n\ge1}\langle g_n,M_n\,g_n\rangle,\quad F_{\mathrm{SLD}}(\Delta)=\sum_{n\ge1}\langle g_n,J_n^{-1}g_n\rangle,
\label{eq:structure}
\end{equation}
and in the order $n$ they are given by
\begin{equation}
M_n=\mathrm{Sym}^n(\Lambda),\quad J_n=\tfrac12\big[\mathrm{Sym}^n(G_+)+\mathrm{Sym}^n(G_-)\big].
\label{eq:blocks}
\end{equation}
For a setting drawn at random and recorded, $M_n$ is the average of $\mathrm{Sym}^n(\Lambda)$ over the settings. The stationary values of the efficiency in the order $n$ are the eigenvalues of the efficiency matrix
\begin{equation}
E_n=J_n^{1/2}M_nJ_n^{1/2},
\label{eq:effmatrix}
\end{equation}
on the range of $J_n$ (the span of its eigenvectors with nonzero eigenvalues). The efficiency of a direction with several orders is the average of the efficiencies of its orders weighted by their quantum Fisher information.
\end{theorem}

The theorem states that a Gaussian measurement and the quantum Fisher information treat every order separately. In every order they are fixed by matrices of the size of phase space, namely by the matrix $\Lambda$, which describes the estimation of a displacement ($M_1=\Lambda$ and $J_1=\openone$), and by the matrices $G_\pm$. In the order $n$ the matrix $\Lambda$ acts on each of the $n$ factors of a mode, like an attenuation that acts independently on each of $n$ photons. The largest and the smallest efficiency in the order $n$ are the extreme eigenvalues of $E_n$. An eigenvector $h_n$ of $E_n$ belongs to the direction with the mode $g_n=J_n^{1/2}h_n$.

We give a self-contained proof in Appendix~\ref{app:structure}, together with the precise conditions. If $g$ is not square integrable, we define the modes $g_n$ by the moments of $\Delta$. The first relation of Eq.~\eqref{eq:structure} holds for every square-integrable $g$, also for directions of infinite $F_{\mathrm{SLD}}$ [Eq.~\eqref{eq:classblocks}]. If a normal mode of the reference state is pure, the matrices $J_n$ with $n\ge2$ have vanishing eigenvalues (all but two of them for a pure state of one mode). Then $F_{\mathrm{SLD}}$ is finite only for the directions whose modes $g_n$ have no component along the eigenvectors of $J_n$ with the eigenvalue zero. The remaining eigenvectors span the range of $J_n$, and $J_n^{-1}$ denotes the inverse on this range.

To see the origin of this structure, we recall that the outcome $\mathbf y$ of a Gaussian measurement is a linear function of a phase-space point $\mathbf x$ with added noise. The ratio $\delta/f$ of the outcome densities at $\mathbf y$ is then the average of $g(\mathbf x)$ over the points $\mathbf x$ that can give this outcome, weighted with their probabilities under $W_0$ (the conditional expectation of $g$ given $\mathbf y$). For normal distributions this average maps a polynomial of degree $n$ in $\mathbf x$ to a polynomial of degree $n$ in $\mathbf y$ and acts on the order $n$ as a symmetric power. It is known that the conditional expectation is the operator called the second quantization of the linear map between the two sets of normal variables~\citep{Janson1997}, and in this way we obtain the classical half of the theorem. We obtain the quantum half from the observation that the product of a Gaussian state with an operator acts on the phase-space function of the operator in the same way, with the matrices $G_\pm$ of the uncertainty relation in place of $\Lambda$. The two signs belong to the two orders of the factors in the product.

A structure of the same kind is known for the correlations of a bipartite Gaussian state with one mode on each side~\citep{BeigiRahimiKeshari2023}. For Gaussian states of finitely many modes without a pure normal mode the blocks of the inner products $\tr(\rho_0^{s}C^\dagger\rho_0^{1-s}Z)$ of two operators $C$ and $Z$ are symmetric powers of the first block for every $0\le s\le1$~\citep{Li2026}. The cases $s=1/2$ and $s=1$ were obtained in Ref.~\cite{LiuWanWu2026}. The symmetrized inner product $\tr[\rho_0(C^\dagger Z+ZC^\dagger)]/2$, in which the quantum Fisher information is the squared norm of the SLD, is the mean of the cases $s=0$ and $s=1$. It is not a member of this family, and its blocks $J_n$ with $n\ge2$ are not symmetric powers of $J_1$. To our knowledge, the comparison of the two halves, i.e., the efficiency of Gaussian measurements in every order, has not been reported. In the first order $J_1=\openone$ and the efficiencies are the eigenvalues of $\Lambda$, which reproduce the known Fisher informations and bounds of the estimation of a displacement~\citep{YuenLax1973,Holevo1982,Genoni2013}.

For one mode $G_\pm=\openone\pm i\Omega/(2v)$ in the normalized quadratures for every $V$. Hence $J_n$ depends only on $\bar n$ and is diagonal in the modes of angular momentum $l=n-2k$, where the index $k=0,\dots,n$ equals the number $k_0$ of radial nodes for $l\ge0$ and $n-k_0$ for $l<0$. Its eigenvalues are
\begin{equation}
\frac{(1+\tfrac1{2v})^{n-k}(1-\tfrac1{2v})^k+(1-\tfrac1{2v})^{n-k}(1+\tfrac1{2v})^k}{2}.
\label{eq:Jeig}
\end{equation}
All dependence on the squeezing of the state and on the measurement is thus contained in the matrices $\Lambda$ of the settings. For $M$ modes $J_n$ is diagonal in the corresponding basis of the normal modes, where its eigenvalues depend only on the occupations $\bar n_j$ and are the means of two products of the factors $1\pm1/(2v_j)$ [Eq.~\eqref{eq:Jdiag}]. For a pure state of $M$ modes it equals $2^{n-1}$ on its range, which has the dimension $2\binom{n+M-1}{n}$.

\subsection{Bounds for all directions}\label{sec:fisher-bounds}
By Theorem~\ref{thm:structure} the search for the direction that a measurement resolves best is reduced to an eigenvalue problem in each order. In the next two theorems we solve this problem for a fixed setting of any number of modes, and for random settings of one mode we bound its solution. We prove them in Appendix~\ref{app:bounds}. The results of Secs.~\ref{sec:fisher-bounds}--\ref{sec:fisher-tmsv} follow from the three theorems, and each of them is derived where we state it. We collect their numerical checks in Appendix~\ref{app:checks}.

\begin{theorem}[one setting]\label{thm:single}
For one Gaussian setting $(H,\Sigma)$ of $M$ modes that obeys Eq.~\eqref{eq:phys} and every direction $\Delta$ with finite $F_{\mathrm{SLD}}$,
\begin{equation}
F(\Delta)\le\lambda_{\max}(\Lambda)\,F_{\mathrm{SLD}}(\Delta),
\label{eq:thmsingle}
\end{equation}
where $\lambda_{\max}(\Lambda)$ is the largest eigenvalue of $HVH^{\mathsf T}(HVH^{\mathsf T}+\Sigma)^{-1}$, which is $V(V+\Sigma)^{-1}$ for general-dyne detection. Equality holds for the displacement of $\rho_0$ along $V^{1/2}\mathbf c$, where $\mathbf c$ is the eigenvector of $\Lambda$ with the largest eigenvalue.
\end{theorem}

Here $\lambda_{\max}(\Lambda)$ is the largest fraction of the variance of the outcomes that comes from the state. No change of the state, Gaussian or not, is thus measured by a fixed Gaussian setting more efficiently than the best displacement. By Sec.~\ref{sec:fisher-chi2} the same bound holds for the $\chi^2$ divergences at any distance from a mixed reference state. It follows that $\lambda_{\max}(\Lambda)$ is the contraction coefficient of the $\chi^2$ divergence under the measurement at $\rho_0$. Among the settings covered by the theorem are general-dyne settings with a mixed seed, e.g., heterodyne detection with inefficient detectors, and measurements of a subset of the modes. For homodyne detection of $d$ commuting quadratures $\lambda_{\max}(\Lambda)=1$ and the bound is the Braunstein--Caves inequality~\citep{BraunsteinCaves1994}. In this case the theorem adds only that the bound is attained in every order, by the $\binom{n+d-1}{n}$ independent directions whose SLDs are polynomials of degree $n$ in the measured quadratures. For one homodyne setting of one mode this is one direction in every order, which in the first order is the displacement along $V\mathbf u$.

For random settings of one mode we write each $\Lambda$ in terms of its eigenvalues $\mu_+\ge\mu_-$ and the phase $\varphi\in[0,\pi)$ of its first eigenvector in the plane of the normalized quadratures. Homodyne detection has $(\mu_+,\mu_-)=(1,0)$, and heterodyne detection near a thermal state $\mu_+=\mu_-=\lambda$. Let
\begin{equation}
\hat w_m=\mathbb E\,e^{2im\varphi},\qquad\omega_n=\max_{1\le m\le n}|\hat w_m|
\label{eq:fourier}
\end{equation}
be the Fourier coefficients of the distribution of the phase and the largest of their moduli up to the order $n$, where $\mathbb E$ denotes the average over the settings. For a fixed phase $|\hat w_m|=1$ for all $m$, and for a uniformly random phase $\hat w_m=0$ for all $m\ge1$. Thus, $\omega_n$ measures how far the distribution of the phase is from the uniform one, as seen by the modes up to the order $n$. Let also $\mu_{\mathrm{iso}}=\mathbb E(\mu_++\mu_-)/2$ and $\mu_{\mathrm{an}}=\mathbb E(\mu_+-\mu_-)/2$ be the mean of the two eigenvalues and half their difference, i.e., the parts of $\Lambda$ that do not and that do depend on the phase.

\begin{theorem}[random settings]\label{thm:mix}
Let the setting of one mode be drawn at random and recorded, let every setting obey Eq.~\eqref{eq:phys}, and let the eigenvalues $\mu_\pm$ be independent of the phase $\varphi$. Then every efficiency in the order $n$ obeys
\begin{equation}
\frac{F}{F_{\mathrm{SLD}}}\le\mu_{\mathrm{iso}}+\omega_n\,\mu_{\mathrm{an}} .
\label{eq:thmmix}
\end{equation}
In the first order the largest efficiency equals $\mu_{\mathrm{iso}}+|\hat w_1|\,\mu_{\mathrm{an}}$ and belongs to a displacement.
\end{theorem}

Theorem~\ref{thm:mix} contains Theorem~\ref{thm:single} for one mode, for which $\omega_n=1$. In addition, it shows what a random phase costs. The isotropic part of $\Lambda$ is kept, while the anisotropic part is reduced by the factor $\omega_n$, which vanishes for a uniform phase. Without the independence of the eigenvalues and the phase, and for any number of modes, the efficiency of a mixture is still bounded by the average of $\lambda_{\max}(\Lambda)$ over the settings. In the first order the largest efficiency is the largest eigenvalue of the average of $\Lambda$. The efficiency of such a mixture is, however, not bounded by that of the best displacement. For example, homodyne detection of $\hat x$ and $\hat p$ with equal probabilities near the vacuum has the efficiency $1/2$ for every displacement and $1$ for the squeezing along $x$ (Appendix~\ref{app:bounds}).

For homodyne detection of one mode with any distribution of the recorded phase, $\mu_{\mathrm{iso}}=\mu_{\mathrm{an}}=1/2$ and the efficiency matrix has an explicit form in the modes of angular momentum $l=n-2k$,
\begin{equation}
\begin{split}
(E_n)_{kk'}&=\sqrt{e_ke_{k'}}\;\hat w_{k'-k},\\
e_k&=\tfrac12\big[\mathbb P(X=k)+\mathbb P(X=n-k)\big],
\end{split}
\label{eq:homkernel}
\end{equation}
where $X$ is binomial with $n$ trials and success probability $\kappa/2=\bar n/(2\bar n+1)$, and $k,k'=0,\dots,n$. Indeed, at the phase $\varphi=0$ the matrix $\Lambda$ projects onto the vector $(1,0)^{\mathsf T}$, which is the sum of the two vectors that belong to the functions $z_1\mp iz_2$, divided by $\sqrt2$. The product of $n$ such factors has the components $2^{-n/2}\binom nk^{1/2}$ along the $n+1$ modes of angular momentum (in the basis of Appendix~\ref{app:conv}), like the amplitudes of $n$ photons each of which is divided equally between two modes. Their squares, multiplied by the eigenvalues~\eqref{eq:Jeig} of $J_n$, are
\begin{equation}
e_k=\frac1{2^{n}}\binom nk\frac{(1+\tau)^{n-k}(1-\tau)^k+(1-\tau)^{n-k}(1+\tau)^k}{2},
\label{eq:ek}
\end{equation}
which is the second relation of Eq.~\eqref{eq:homkernel} for the success probability $(1-\tau)/2=\kappa/2$. The efficiency matrix of this setting has the elements $\sqrt{e_ke_{k'}}$. If we rotate the setting by $\varphi$, they are multiplied by $e^{2i(k'-k)\varphi}$, and after the average over the phase we obtain Eq.~\eqref{eq:homkernel} [see Eq.~\eqref{eq:Emix}].

The numbers $e_k$ sum to one, and none of them exceeds $1/2$. For $k\ne n-k$ this holds because the two probabilities in Eq.~\eqref{eq:homkernel} belong to different outcomes of $X$. For $k=n/2$ it follows from $e_k=\binom n{n/2}[\kappa(2-\kappa)/4]^{n/2}\le2^{-n}\binom n{n/2}\le1/2$. We now show that for homodyne detection of one mode with a recorded phase of any distribution, every efficiency in the order $n$ is at most
\begin{equation}
e^{\ast}+\omega_n(1-e^{\ast})\le\frac{1+\omega_n}{2},\qquad e^{\ast}=\max_ke_k\le\frac12 .
\label{eq:corhom}
\end{equation}
To see this, we separate the diagonal elements of $E_n$ from the others, for which $|\hat w_{k'-k}|\le\omega_n$. For a unit vector $\boldsymbol\upsilon$ with the components $\upsilon_k$,
\begin{equation}
\begin{split}
\langle\boldsymbol\upsilon,E_n\boldsymbol\upsilon\rangle&\le(1-\omega_n)\sum_ke_k|\upsilon_k|^2+\omega_n\Big(\sum_k\sqrt{e_k}\,|\upsilon_k|\Big)^2\\
&\le(1-\omega_n)e^\ast+\omega_n,
\end{split}
\label{eq:homsplit}
\end{equation}
where the second step is the Cauchy--Schwarz inequality.

We state the bounds~\eqref{eq:thmmix} and~\eqref{eq:corhom} for each order because $\omega_n$ grows with $n$. For a finite set of phases $\omega_n$ tends to one, so that the bound becomes trivial at high orders. Indeed, by Dirichlet's theorem on simultaneous approximation there are integers $m\ge1$ for which $2m\varphi$ is arbitrarily close to a multiple of $2\pi$ for all phases of the set at once. If the differences of the phases are commensurate with $\pi$, $\omega_n$ equals one from some order on. For $K$ equally spaced phases near a thermal state we obtain the exact value instead. For the phases $\varphi_{\mathrm{off}}+j\pi/K$ with $j=0,\dots,K-1$ and an offset $\varphi_{\mathrm{off}}$ we have $\hat w_m=e^{2im\varphi_{\mathrm{off}}}$ if $K$ divides $m$ and $\hat w_m=0$ otherwise. The matrix~\eqref{eq:homkernel} then consists of $K$ separate blocks, one for each remainder $a$ of the division of $k$ by $K$. The block $a$ has the elements $\sqrt{e_ke_{k'}}e^{2i(k'-k)\varphi_{\mathrm{off}}}$. This block is the outer product of one vector with itself, and its only nonzero eigenvalue is the sum $\sum_{k\equiv a}e_k$ over its indices. Hence the largest efficiency in the order $n$ is $\max_a\sum_{k\equiv a\,(\mathrm{mod}\,K)}e_k$, which is the uniform-phase value $e^\ast$ for $n<K$. Since the matrix couples only modes whose angular momenta differ by multiples of $2K$, for a direction with $|l|<K$ in all its modes the Fisher information equals that of a uniform phase.

\subsection{Thermal states}\label{sec:fisher-thermal}
Near a thermal state of one mode, $V=v\,\openone$, both quadrature detections commute with the rotations of phase space. In this case we obtain every efficiency in closed form from Theorem~\ref{thm:structure}. Let $\bar n>0$ and let $b_{n,l}$ be the coefficients of $g$ in the orthonormal modes of order $n$ and angular momentum $l$.

For heterodyne detection we have $\Lambda=\lambda\openone$ and, consequently, $M_n=\lambda^n$. The efficiency matrix $E_n=\lambda^nJ_n$ is then diagonal in the modes of angular momentum. Since $\lambda(1+\tau)=1$ and $\lambda(1-\tau)=q$, we obtain from the eigenvalues~\eqref{eq:Jeig} of $J_n$ the efficiency $(q^k+q^{n-k})/2$ for the mode with $l=n-2k$. For homodyne detection at a uniformly random phase $\hat w_m=0$ for $m\ge1$, so $E_n=\mathrm{diag}(e_k)$ by Eq.~\eqref{eq:homkernel}, and $M_n$ is diagonal with the elements $2^{-n}\binom nk$. Both sets of efficiencies are symmetric under the exchange of $k$ and $n-k$, i.e., of $l$ and $-l$. We write them in terms of the number of radial nodes, $k_0=(n-|l|)/2=\min(k,n-k)$. For this number the two indices $k$ and $n-k$ are $k_0$ and $k_0+|l|$, and the efficiencies of the mode $(n,l)$ read
\begin{equation}
\begin{split}
\frac{F_Q}{F_{\mathrm{SLD}}}&=\frac{\mathbb P(N_0\ge k_0)+\mathbb P(N_0\ge k_0+|l|)}{2}=\frac{1+q^{|l|}}{2}\,q^{k_0},\\
\frac{F_W}{F_{\mathrm{SLD}}}&=\frac{\mathbb P(X=k_0)+\mathbb P(X=k_0+|l|)}{2},
\end{split}
\label{eq:spectrum}
\end{equation}
where the random variable $N_0$ (not a number of events) is the photon number of $\rho_0$, with $\mathbb P(N_0\ge m)=q^m$, and $X$ is the binomial variable of Eq.~\eqref{eq:homkernel}. The Fisher informations of the two detections are
\begin{equation}
F_Q=\sum_{n,l}\lambda^n|b_{n,l}|^2,\qquad F_W=\sum_{n,l}\frac{1}{2^n}\binom{n}{k_0}|b_{n,l}|^2,
\label{eq:modes}
\end{equation}
and dividing either of them mode by mode by its efficiency gives the quantum Fisher information,
\begin{equation}
F_{\mathrm{SLD}}=\sum_{n,l}\frac{2\lambda^n}{(1+q^{|l|})\,q^{k_0}}\,|b_{n,l}|^2 .
\label{eq:sldmodes}
\end{equation}
For every direction, the bound~\eqref{eq:corhom} with $\omega_n=0$ and Theorem~\ref{thm:single} with $\lambda_{\max}(\Lambda)=\lambda$ give
\begin{equation}
F_W(\Delta)\le\tfrac12F_{\mathrm{SLD}}(\Delta),\qquad F_Q(\Delta)\le\lambda\,F_{\mathrm{SLD}}(\Delta),
\label{eq:thermalbounds}
\end{equation}
with equality for displacements. With the second bound we sharpen the Husimi side of the bounds of Ref.~\cite{Chitgopekar2026}, $F_Q\le F_{\mathrm{SLD}}$ for unitary signals, to $F_Q\le\lambda F_{\mathrm{SLD}}$ for all directions.

\begin{figure}[t]
\centering\includegraphics[width=\columnwidth]{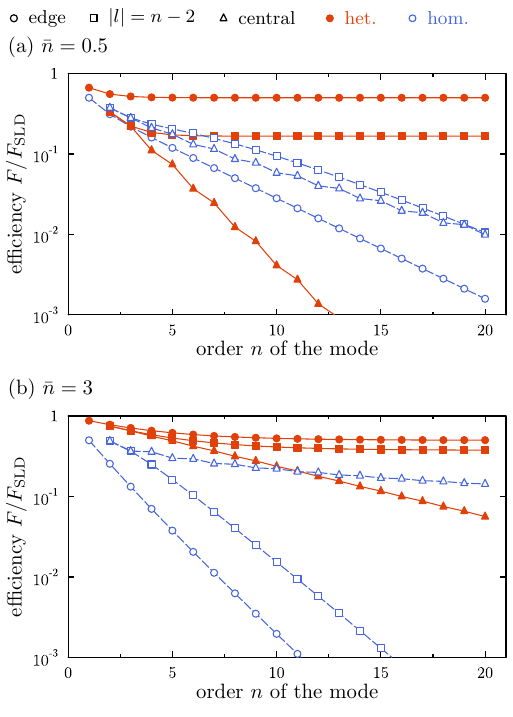}
\caption{Efficiency $F/F_{\mathrm{SLD}}$ of the Hermite--Gauss modes of order $n$ and angular momentum $l$ near a thermal state, Eq.~\eqref{eq:spectrum}, at (a) $\bar n=0.5$ and (b) $\bar n=3$. Filled markers and solid lines show heterodyne detection (het.), and open markers with dashed lines homodyne detection at a uniformly random phase (hom.). Circles denote edge modes ($|l|=n$), squares the modes with $|l|=n-2$ and triangles central modes ($|l|\le1$). Efficiencies below $10^{-3}$ are not shown.}\label{fig:modeeff}
\end{figure}

Figure~\ref{fig:overview}(b) shows the efficiencies~\eqref{eq:spectrum} of all modes at $\bar n=1$, and in Fig.~\ref{fig:modeeff} we plot their dependence on the order at $\bar n=0.5$ and $\bar n=3$. In heterodyne detection a mode of order $n$ is damped by the factor $\lambda^n$. This is the Gaussian smoothing that turns $W$ into $Q$, seen in the basis of $W_0$. In homodyne detection at a uniformly random phase, the $n+1$ angular components of order $n$ share unit weight according to the probabilities $2^{-n}\binom nk$ of $k$ successes in $n$ fair trials. These probabilities are the squared singular values of the Radon transform with a Gaussian weight~\citep{Davison1981,Louis1985}. To these known weights we add the quantum Fisher information of the same modes, Eq.~\eqref{eq:sldmodes}, and with it the efficiencies. In the homodyne efficiency the fair trials become trials with success probability $\kappa/2$. The two outcomes $k_0$ and $k_0+|l|$ (they belong to the angular momenta $\pm l$) enter it with equal weights. At large occupation we have $\kappa/2\to1/2$ and $q\to1$. Heterodyne detection then reaches the quantum limit in every mode, whereas homodyne detection keeps only the angular weight of the mode. Near the vacuum both efficiencies tend to $1/2$ for the edge modes ($k_0=0$) and to zero for the others, whose quantum Fisher information diverges.

The quotient of the homodyne and the heterodyne efficiency of one mode is $\binom n{k_0}(1-\kappa/2)^n$. Hence, heterodyne detection gives more information on the mode $(n,l)$ if and only if
\begin{equation}
\binom{n}{\frac{n-|l|}{2}}<\left(\frac{2\bar n+1}{\bar n+1}\right)^n .
\label{eq:criterion}
\end{equation}
This holds for every edge mode when $\bar n>0$, and for none of the central modes of order $n\ge4$, $8$ and $20$ at $\bar n=1$, $2$ and $5$. The mode $(2,0)$, which describes a change of the temperature, favors heterodyne detection if and only if $\bar n>1/\sqrt2$. This is the switch between homodyne and heterodyne detection as the optimal Gaussian thermometer found in Ref.~\cite{Cenni2022}. The heterodyne efficiency $q$ of this mode agrees with the Fisher informations derived there and with the ratio given in Ref.~\cite{Kerskens2026}. With Eq.~\eqref{eq:criterion} we can reconcile the known result that heterodyne detection is better for displacements and, for almost all Gaussian states, for quadrature moments up to second order~\citep{YuenLax1973,Muller2016,Teo2017} with the known result that homodyne detection is better in full tomography~\citep{Fernandes2025}. The quadrature moments up to second order are modes with $n\le2$, whereas in full tomography the figure of merit weighs the fine structure of high order.

Among the modes with a given angular momentum $|l|\ge1$ the largest heterodyne efficiency is $(1+q^{|l|})/2>1/2$. For every $|l|\ge1$ the best heterodyne direction is therefore measured more efficiently than the best homodyne direction. For typical directions that change the non-Gaussian structure the efficiencies are much smaller than the bounds. For each Fock support and each occupation we have drawn $200$ random directions, with independent normal coefficients in a basis of the traceless Hermitian matrices. The Fock support extended up to photon number $3$, $5$ or $7$, and the occupation $\bar n$ was taken between $0.5$ and $5$. For both detections the median efficiencies over these directions are 0.04\mbox{--}0.19. These medians depend on the ensemble, and they fall as the support grows, from $0.14$--$0.19$ for photon numbers up to $3$ to $0.04$--$0.10$ up to $7$.

\subsection{Coherent and squeezed states}\label{sec:fisher-squeezed}
The closed forms of Sec.~\ref{sec:fisher-thermal} do not cover two kinds of reference states of one mode. We treat them in turn. For coherent states the matrices $J_n$ with $n\ge2$ are singular. In the case of squeezed thermal states it is the rotation symmetry of phase space that is lost.

Near a coherent state $\rho_0=\ket\alpha\!\bra\alpha$ heterodyne detection and homodyne detection at a uniform phase have the efficiency $1/2$ in every direction of finite $F_{\mathrm{SLD}}$, as we now show. The directions with finite $F_{\mathrm{SLD}}$ are $\Delta=\ket\psi\!\bra\alpha+\ket\alpha\!\bra\psi$ with $\braket{\alpha|\psi}=0$ (Appendix~\ref{app:structure}). In the first order of $\varepsilon$ they change the coherent state into another pure state, as a small unitary operation does. An incoherent admixture, e.g., of thermal noise, has instead an infinite $F_{\mathrm{SLD}}$. By Eq.~\eqref{eq:Jeig} with $\tau=1$, the range of $J_n$ is spanned by the two edge modes, $l=\pm n$ (the indices $k=0$ and $k=n$ of Eq.~\eqref{eq:Jeig}), and $J_n=2^{n-1}$ on it. With $\Lambda=\openone/2$ for heterodyne detection, $E_n=2^{-n}J_n$ equals $1/2$ on the range, i.e., heterodyne detection has the efficiency $1/2$ in every direction of finite $F_{\mathrm{SLD}}$. For homodyne detection $e_0=e_n=1/2$ and all other $e_k$ vanish. By Eq.~\eqref{eq:homkernel}, $E_n$ on the range is then $\tfrac12\big(\begin{smallmatrix}1&\hat w_n\\\overline{\hat w_n}&1\end{smallmatrix}\big)$, so homodyne detection has the efficiencies $(1\pm|\hat w_n|)/2$ in the order $n$. They equal $1/2$ for a uniform phase. For $K$ equally spaced phases $|\hat w_n|=1$ if $K$ divides $n$ and $\hat w_n=0$ otherwise. In this case the efficiency is $1/2$ in the orders that are not multiples of $K$, whereas in the orders that are multiples of $K$ the efficiencies are $1$ and $0$.

The value $1/2$ is the largest that a measurement can have on average over the directions. Since the Wigner function of $\ket n\!\bra0$ is proportional to $(x-ip)^ne^{-x^2-p^2}$, the edge modes of order $n$ in the Fock basis displaced by $\alpha$ are the directions $\ket n\!\bra0+\mathrm{h.c.}$ and $i\ket n\!\bra0+\mathrm{h.c.}$, which include the displacement ($n=1$) and the squeezing ($n=2$). No measurement can have a larger average efficiency than $1/2$ over the edge modes of the orders $n\le D$, for any integer $D\ge1$. To show this, let $P_D$ project onto the span of the displaced Fock states $\ket0,\dots,\ket D$. The operators of any measurement of the oscillator, multiplied by $P_D$ from both sides, form a measurement in dimension $D+1$ with the same outcome distributions for the states in this span. By the Gill--Massar bound~\citep{GillMassar2000}, for a pure state in dimension $D+1$ the efficiencies of any measurement in $2D$ directions that are orthogonal in the inner product of the SLDs, such as these edge modes, add up to at most $D$. Their average is therefore at most $1/2$. A measurement that has the efficiency $1/2$ in every direction, and thus divides this budget equally, is called Fisher-symmetric~\citep{LiFerrie2016,ZhuHayashi2018}. Since both quadrature detections reach $1/2$ in every direction and for every $D$, they are Fisher-symmetric at coherent states, in infinite dimension. A measurement that does better in one direction, such as homodyne detection at the phase of a displacement, which reaches $F_{\mathrm{SLD}}$, must do worse in others. Near a coherent state Eq.~\eqref{eq:modes} holds with $\lambda=1/2$. Since no mode then favors heterodyne detection, we obtain $F_W\ge F_Q$ for every direction, with equality for the directions of finite $F_{\mathrm{SLD}}$.

Let now $\rho_0$ be a thermal state of occupation $\bar n$ squeezed by $r$, possibly displaced, with $V=vR_\chi\,\mathrm{diag}(e^{-2r},e^{2r})R_\chi^{\mathsf T}$, where $R_\chi$ is the rotation by the angle $\chi$ of the quadrature with the variance $ve^{-2r}$. For heterodyne detection $\Lambda$ has the eigenvalues $ve^{\mp2r}/(ve^{\mp2r}+\tfrac12)$. Using Theorem~\ref{thm:single} with the larger of them we obtain, for every direction,
\begin{equation}
F_Q\le\lambda_r\,F_{\mathrm{SLD}},\qquad\lambda_r=\frac{e^{2|r|}(2\bar n+1)}{e^{2|r|}(2\bar n+1)+1},
\label{eq:sqhet}
\end{equation}
with equality for the displacement along the antisqueezed quadrature, the eigenvector of $V$ with the eigenvalue $ve^{2|r|}$. For homodyne detection at the laboratory phase $\theta$ the vector $V^{1/2}\mathbf u$ has the phase $\varphi$ in the plane of the normalized quadratures, with $\tan(\varphi-\chi)=e^{2r}\tan(\theta-\chi)$. A uniform phase of the local oscillator is therefore not uniform in the normalized quadratures. If $\theta$ is uniform, the density of $\varphi$ with respect to $d\varphi/\pi$ is
\begin{equation}
\begin{split}
\frac{d\theta}{d\varphi}&=\frac{1}{\cosh2r+\sinh2r\cos2(\varphi-\chi)}\\
&=\frac{1-\tanh^2r}{1+\tanh^2r+2\tanh r\cos2(\varphi-\chi)},
\end{split}
\label{eq:poisson}
\end{equation}
the Poisson kernel, whose Fourier coefficients are $\hat w_m=(-\tanh r)^{|m|}e^{2im\chi}$, i.e., the phases near the antisqueezed quadrature are sampled more often in the units of the state. Hence $\omega_n=\tanh|r|$ for all $n$, and Eq.~\eqref{eq:corhom} gives
\begin{equation}
F_W\le\frac{1+\tanh|r|}{2}\,F_{\mathrm{SLD}}
\label{eq:sqhom}
\end{equation}
for every direction. By Theorem~\ref{thm:mix} this bound is attained in the first order. Since the eigenvector of $\mathbb E\Lambda$ with the largest eigenvalue is then along the antisqueezed quadrature, the bound is again attained by the displacement along this quadrature.

\begin{figure}[t]
\centering\includegraphics[width=\columnwidth]{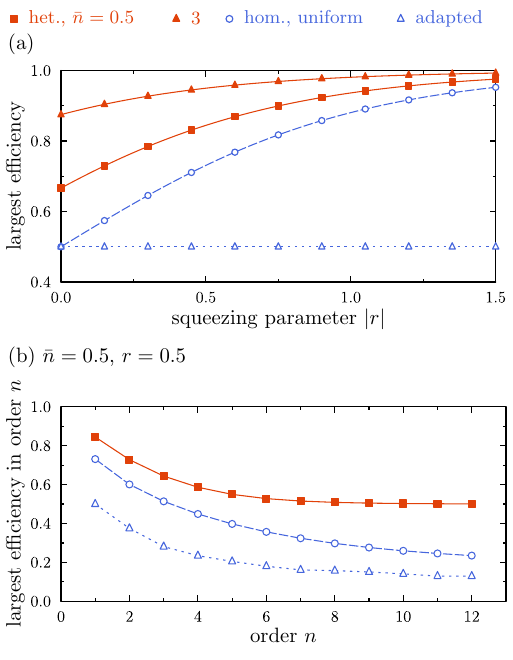}
\caption{Efficiencies near a squeezed thermal state. (a) Largest efficiency over all directions against the squeezing parameter $|r|$. Filled markers and solid lines show heterodyne detection, Eq.~\eqref{eq:sqhet}, for the occupations $\bar n=0.5$ (squares) and $3$ (triangles). Open circles and the dashed line show homodyne detection at a uniformly random phase, Eq.~\eqref{eq:sqhom}. This curve does not depend on $\bar n$ and coincides with the heterodyne bound for $\bar n=0$. Open triangles, dotted line: homodyne detection at a random phase that is uniform in the normalized quadratures (``adapted'' in the legend). Panel (b) shows the largest efficiency in each order $n$ at $\bar n=0.5$ and $r=0.5$ for heterodyne detection (squares) and the two homodyne detections (circles and triangles as in the upper panel).}\label{fig:squeezed}
\end{figure}

Both bounds are thus raised toward one when the reference state is squeezed [Figs.~\ref{fig:overview}(c) and~\ref{fig:squeezed}(a)]. The reason is that the antisqueezed quadrature becomes noisy, and the noise added by the measurement matters less. For a squeezed vacuum the bounds~\eqref{eq:sqhet} and~\eqref{eq:sqhom} coincide. Under a uniform laboratory phase the modes of one order are no longer measured independently. The efficiencies of order $n$ are the eigenvalues of the matrix~\eqref{eq:homkernel}, and the largest of them decreases with the order [Fig.~\ref{fig:squeezed}(b)]. If the phase is instead distributed uniformly in the normalized quadratures, $E_n=\mathrm{diag}(e_k)$ as for the thermal state. In this way we recover the binomial spectrum of Eq.~\eqref{eq:spectrum}, but at the cost of the largest efficiency, which falls back to $1/2$.

\subsection{Pure states of several modes}\label{sec:fisher-pure}
For several modes we obtain the efficiencies of local and of joint Gaussian measurements from Theorems~\ref{thm:structure} and~\ref{thm:single}. Here we evaluate them for pure states, and in Sec.~\ref{sec:fisher-tmsv} for the two-mode squeezed vacuum. A pure Gaussian state of $M$ modes is $\ket{\Psi_0}=U\ket{0}$ with a Gaussian unitary operation $U$ and the vacuum $\ket0$ of all modes. Let $S$ be the symplectic matrix of $U$, $U^\dagger\hat{\mathbf x}U=S\hat{\mathbf x}$, so that $V=SS^{\mathsf T}/2$, and let $r_1,\dots,r_M$ be the squeezing parameters of $\ket{\Psi_0}$, for which the eigenvalues of $2V$ are $e^{\pm2r_j}$. As for one mode, the directions of finite $F_{\mathrm{SLD}}$ are $\Delta=\ket{\Psi_0}\!\bra\psi+\ket\psi\!\bra{\Psi_0}$ with $\braket{\Psi_0|\psi}=0$. We expand $\ket\psi=\sum_{\mathbf m}\psi_{\mathbf m}U\ket{\mathbf m}$ in the images of the Fock states $\ket{\mathbf m}=\ket{m_1,\dots,m_M}$.

We first describe the range of $J_n$ at a pure state. It consists of the modes built from the creation operators only or from the annihilation operators only of the modes that annihilate $\ket{\Psi_0}$. In this section and the next we normalize the quadratures with the matrix $\Gamma=S/\sqrt2$ in place of $V^{1/2}$, $\mathbf z=\Gamma^{-1}\mathbf x$. Theorem~\ref{thm:structure} remains unchanged under this choice (Appendix~\ref{app:conv}). Up to the factor $\sqrt2$ these coordinates are the quadratures of the Bogoliubov modes $\hat a_j'=U\hat a_jU^\dagger$ that annihilate $\ket{\Psi_0}$. In this way the reference state is described as the vacuum of these modes. Here $G_\pm=\openone\pm i\Omega$. The matrix $i\Omega$ has the eigenvalues $\pm1$, and $G_\pm/2$ are the projectors onto the two spaces $\mathcal W_\pm$ of its eigenvectors. The space $\mathcal W_+$ has the orthonormal basis $\mathbf e_+^{(j)}$, $j=1,\dots,M$, with the entries $(1,-i)/\sqrt2$ in the block of the mode $j$, and $\mathcal W_-$ has the basis $\mathbf e_-^{(j)}=\overline{\mathbf e_+^{(j)}}$. These are the vectors $\mathbf e_\pm^{(j)}$ of Appendix~\ref{app:conv}, where the orthogonal matrix $O$ defined there equals $\openone$ in the present coordinates. The linear functions $\mathbf e_\pm^{(j)}\cdot\mathbf z/\sqrt2$ are the phase-space functions of the operators $\hat a_j'^\dagger$ and $\hat a_j'$, respectively. We denote by $\Sym^n(\mathcal W_+)$ the modes of order $n$ that are built from the functions $\mathbf e_+^{(j)}\cdot\mathbf z$ only, which correspond to products of $n$ creation operators, and by $\Sym^n(\mathcal W_-)$ those built from the functions $\mathbf e_-^{(j)}\cdot\mathbf z$ only. Since $G_\pm/2$ are projectors, $J_n=2^{n-1}[\Sym^n(G_+/2)+\Sym^n(G_-/2)]$ is $2^{n-1}$ times the projector $\Pi_n$ onto the modes that are sums of a vector of $\Sym^n(\mathcal W_+)$ and a vector of $\Sym^n(\mathcal W_-)$. The range of $J_n$ thus contains no mode with products of creation and annihilation operators of the modes $\hat a_j'$.

\emph{Modes of a direction.} The order $n$ of a direction consists of the components of $\ket\psi$ with $n$ photons in total, $|\mathbf m|=m_1+\dots+m_M=n$. To show this and to find the mode $g_n$, let $\ket\psi=\sum_{|\mathbf m|=n}\psi_{\mathbf m}U\ket{\mathbf m}$. With the creation operators $\hat a_j'^\dagger=U\hat a_j^\dagger U^\dagger$, the operator
\begin{equation}
A=\sum_{\mathbf m}\psi_{\mathbf m}\prod_j\frac{(\hat a_j'^\dagger)^{m_j}}{\sqrt{m_j!}}
\label{eq:Apoly}
\end{equation}
is a polynomial in the quadratures, $A\ket{\Psi_0}=\ket\psi$ and $\bra{\Psi_0}A=0$. The operator $2(A+A^\dagger)$ therefore agrees with the SLD $2\Delta$ in the row and the column of $\ket{\Psi_0}$. Since for a pure state these are the only elements that enter Eq.~\eqref{eq:fsld}, $2(A+A^\dagger)$ can be used as the SLD of $\Delta$ (Appendix~\ref{app:structure}). We now go over to phase space, where an operator is represented by a function in the same way as $\rho$ is represented by $W$ (Appendix~\ref{app:conv}). The function of $\hat a_j'^\dagger$ is $\mathbf e_+^{(j)}\cdot\mathbf z/\sqrt2$. The function of a product of such operators is the product of their functions, without correction terms. This holds because the operators $\hat a_j'^\dagger$ commute with each other [formally, $\mathbf e_+^{(j)}\cdot\mathbf e_+^{(j')}=0$ in Eq.~\eqref{eq:genf}]. Hence the function of $A$ is a mode of order $n$ and corresponds to the vector $2^{-n/2}\hat\psi$ of $\Sym^n(\mathcal W_+)$. Here $\hat\psi=\sum_{\mathbf m}\psi_{\mathbf m}\ket{\mathbf m}_+$ and $\ket{\mathbf m}_+$ is the normalized symmetrization of the product of $m_j$ factors $\mathbf e_+^{(j)}$ for every $j$ (the state $\ket{\mathbf m,0}$ of Appendix~\ref{app:conv}). We obtain the mode of a direction by applying $J_n$ to the phase-space function of its SLD [Eq.~\eqref{eq:quantblocks}]. With $J_n=2^{n-1}\Pi_n$ this gives
\begin{equation}
g_n=2^{n/2}\big(\hat\psi+\overline{\hat\psi}\big),\qquad F_{\mathrm{SLD}}(\Delta)=4\braket{\psi|\psi} .
\label{eq:puremodes}
\end{equation}
These $g_n$ exhaust the real vectors in the range of $J_n$, and a general $\ket\psi$ is the sum of its components with $n=1,2,\dots$ photons.

\emph{Pairs of efficiencies.} For a setting that measures all modes and adds no more noise than the uncertainty relation requires, the efficiencies of every order come in pairs with the mean $1/2$, as we now show. Let a setting obey, in these coordinates,
\begin{equation}
\Omega\Lambda\Omega^{\mathsf T}=\openone-\Lambda .
\label{eq:complete}
\end{equation}
In such a setting all modes are measured and no noise is added beyond that required by the uncertainty relation. The condition holds for general-dyne detection of all modes with a pure seed. For this detection $\Lambda=(\openone+\Sigma_S)^{-1}$ with the symmetric, positive and symplectic matrix $\Sigma_S=2S^{-1}\Sigma(S^{\mathsf T})^{-1}$ and $\Omega\Sigma_S\Omega^{\mathsf T}=\Sigma_S^{-1}$. Homodyne detection of $M$ commuting quadratures obeys the condition too. In this case $\Lambda$ projects onto a space of dimension $M$ that $\Omega$ maps onto its orthogonal complement (Appendix~\ref{app:bounds}). The same is true for limits of pure seeds, such as homodyne detection of some modes and general-dyne detection of the others. Let $\Theta=\openone-2\Lambda$. By Eq.~\eqref{eq:complete}, $\Omega\Theta=-\Theta\Omega$, so $\Theta$ maps $\mathcal W_\mp$ into $\mathcal W_\pm$. We denote by $\Phi$ the restriction of $-\Theta$ to $\mathcal W_-$, a map from $\mathcal W_-$ to $\mathcal W_+$ with the largest singular value $\|\Phi\|\le1$. Then $G_+\Lambda G_+=G_+$ and $G_+\Lambda G_-=2\Phi$ on $\mathcal W_-$, and $E_n=2^{n-1}\Pi_n\Sym^n(\Lambda)\Pi_n$. For a recorded mixture of such settings we obtain, in the decomposition $\Sym^n(\mathcal W_+)\oplus\Sym^n(\mathcal W_-)$,
\begin{equation}
E_n=\frac12\begin{pmatrix}\openone&\Phi_n\\\Phi_n^\dagger&\openone\end{pmatrix},\qquad\Phi_n=\mathbb E\,\Sym^n(\Phi).
\label{eq:pair}
\end{equation}
The eigenvalues of this matrix are $(1\pm\mathsf s)/2$, where $\mathsf s$ runs over the singular values of $\Phi_n$, the square roots of the eigenvalues of $\Phi_n^\dagger\Phi_n$. Thus, for general-dyne detection of all modes with a pure seed, for homodyne detection of $M$ commuting quadratures, and for recorded mixtures of such settings, the efficiencies of every order come in pairs $(1\pm\mathsf s)/2$ with $0\le\mathsf s\le1$. Their mean is $1/2$ in every order. For the direction of Eq.~\eqref{eq:puremodes},
\begin{equation}
\frac{F}{F_{\mathrm{SLD}}}=\frac12+\frac{\mathrm{Re}\,\langle\hat\psi,\Phi_n\overline{\hat\psi}\rangle}{2\braket{\psi|\psi}} .
\label{eq:effpsi}
\end{equation}
All these measurements have the mean efficiency of the Fisher-symmetric detections of Sec.~\ref{sec:fisher-squeezed}, but in general they distribute the efficiency unevenly over the directions. General-dyne detection whose seed is the reference state itself, $\Sigma=V$, has $\Lambda=\openone/2$ and $\Phi=0$. It is Fisher-symmetric at every pure Gaussian state, as heterodyne detection is at coherent states.

\emph{Heterodyne detection.} For $\Sigma=\openone/2$ the matrix $\Sigma_S=(S^{\mathsf T}S)^{-1}$ has the eigenvalues $e^{\mp2r_j}$, and $\Theta=(\Sigma_S-\openone)(\Sigma_S+\openone)^{-1}$ the eigenvalues $\mp\tanh r_j$. Since $\Phi^\dagger\Phi$ is the restriction of $\Theta^2$ to $\mathcal W_-$ and $\Theta^2$ commutes with $\Omega$, the singular values of $\Phi$ are $\tanh|r_j|$, and those of $\Sym^n(\Phi)$ are their products of degree $n$. The efficiencies of heterodyne detection of every mode in the order $n$ are therefore
\begin{equation}
\frac12\Big(1\pm\prod_{j=1}^M\tanh^{n_j}|r_j|\Big),\qquad n_1+\dots+n_M=n .
\label{eq:purespec}
\end{equation}
Here the integer $n_j\ge0$ is the number of factors $\tanh|r_j|$ in the product. The largest of these efficiencies belongs to the first order, where it is attained by a displacement (Theorem~\ref{thm:single}), so that
\begin{equation}
F_Q\le\tfrac12\big(1+\max_j\tanh|r_j|\big)F_{\mathrm{SLD}}
\label{eq:purebound}
\end{equation}
for every direction.

\emph{Homodyne detection of every mode.} Homodyne detection of one quadrature of every mode at uniformly random phases has the Fisher information of heterodyne detection of every mode, $F_W=F_Q$, in every direction of finite $F_{\mathrm{SLD}}$. To show this, we compare local homodyne detection with heterodyne detection through the dependence of $\Phi$ on the seed. For a pure seed let $\mathcal K(\Sigma_S)$ be the space of the vectors $\mathbf v\in\mathbb C^{2M}$ with $\Sigma_S\mathbf v=-i\Omega\mathbf v$. Since $\Sigma_S=(\openone+\Theta)(\openone-\Theta)^{-1}$,
\begin{equation}
\mathcal K(\Sigma_S)=\{\boldsymbol\beta+\Phi\boldsymbol\beta:\ \boldsymbol\beta\in\mathcal W_-\},
\label{eq:graph}
\end{equation}
because $\mathbf v=(\openone-\Theta)\boldsymbol\beta$ obeys $\Sigma_S\mathbf v=(\openone+\Theta)\boldsymbol\beta=-i\Omega\mathbf v$ for $\boldsymbol\beta\in\mathcal W_-$, and both spaces have the dimension $M$. Let $\varsigma$ be the map $\Phi$ of the same seed for the vacuum as the reference state ($S=\openone$), for which $\mathcal K(2\Sigma)$ is the set of the vectors $\boldsymbol\beta+\varsigma\boldsymbol\beta$. The relation $S^{-1}\Omega=\Omega S^{\mathsf T}$ gives $\mathcal K(\Sigma_S)=S^{\mathsf T}\mathcal K(2\Sigma)$. In the basis $\mathbf e_+^{(j)},\mathbf e_-^{(j)}$ the real matrix $S^{\mathsf T}$ has the blocks $\big(\begin{smallmatrix}\mathsf a&\mathsf b\\\bar{\mathsf b}&\bar{\mathsf a}\end{smallmatrix}\big)$. Applying them to the vectors $\boldsymbol\beta+\varsigma\boldsymbol\beta$ and comparing with Eq.~\eqref{eq:graph} gives
\begin{equation}
\Phi=(\mathsf a\varsigma+\mathsf b)(\bar{\mathsf b}\varsigma+\bar{\mathsf a})^{-1}.
\label{eq:mobius}
\end{equation}
Since $\mathsf a\mathsf a^\dagger-\mathsf b\mathsf b^\dagger=\openone$ for a symplectic matrix, we have $\|\bar{\mathsf a}^{-1}\bar{\mathsf b}\|<1$, and the inverse exists for $\|\varsigma\|\le1$. Every element of $\Phi$ is thus an analytic function of the elements of $\varsigma$ for $\|\varsigma\|<1$ and a continuous one for $\|\varsigma\|\le1$. Heterodyne detection has $\varsigma=0$, and homodyne detection of the quadratures at the phases $\theta_j$ is the limit $\varsigma=\mathrm{diag}(e^{2i\theta_j})$. Since the real matrix $\Theta$ is fixed by $\Phi$, in this limit $\Lambda=(\openone-\Theta)/2$ tends to the projector of the homodyne setting. A shift of all phases by the same angle $\theta'$ multiplies $\varsigma$ by $e^{2i\theta'}$. We therefore consider $\Sym^n[\Phi(\zeta\varsigma)]$ as a function of a complex number $\zeta$. It is analytic for $|\zeta|<1$ and continuous for $|\zeta|\le1$. For such a function the mean over the circle $|\zeta|=1$ equals its value at the center $\zeta=0$ (the mean-value property of analytic functions). This value is $\Sym^n(\Phi)$ of heterodyne detection. Consequently, if the joint distribution of the phases does not change when all phases are shifted by the same angle, $\Phi_n$ is that of heterodyne detection. By Eq.~\eqref{eq:effpsi}, homodyne detection of one quadrature of every mode at such phases has the Fisher information $F_Q$ of heterodyne detection of every mode for every direction of finite $F_{\mathrm{SLD}}$. In particular, $F_W=F_Q$ for these directions. The statement covers independent uniform phases and also one uniformly random phase common to all local oscillators, whatever the fixed differences of their phases are. Note that at a pure state no direction of finite $F_{\mathrm{SLD}}$ distinguishes heterodyne detection from homodyne detection at uniform phases. The two detections differ for directions of infinite $F_{\mathrm{SLD}}$ and at mixed reference states.

\subsection{Two-mode squeezed vacuum}\label{sec:fisher-tmsv}
As an example of two modes $a$ and $b$ we take as the reference state $\ket{\Psi_0}$ the two-mode squeezed vacuum,
\begin{equation}
\ket{\Psi_r}=e^{r(\hat a^\dagger\hat b^\dagger-\hat a\hat b)}\ket{0,0}=\frac{1}{\cosh r}\sum_{m\ge0}\tanh^mr\,\ket{m,m},
\label{eq:tmsv}
\end{equation}
with $r>0$. Each arm of this state is in a thermal state with the occupation $\bar n=\sinh^2r$. Here the vectors $U\ket{m_a,m_b}$ are the Fock states of the Bogoliubov modes $\hat a\cosh r-\hat b^\dagger\sinh r$ and $\hat b\cosh r-\hat a^\dagger\sinh r$, which annihilate $\ket{\Psi_r}$. We write $\psi_m$ for the component of a direction of the order $n$ along $U\ket{m,n-m}$. The symplectic matrix is
\begin{equation}
S=\begin{pmatrix}\cosh r\,\openone&\sinh r\,\sigma_z\\\sinh r\,\sigma_z&\cosh r\,\openone\end{pmatrix}
\label{eq:Stmsv}
\end{equation}
with the Pauli matrices $\sigma_x$ and $\sigma_z$. In the basis $\mathbf e_+^{(a)},\mathbf e_+^{(b)}$ we then have $\mathsf a=\cosh r\,\openone$ and $\mathsf b=\sinh r\,\sigma_x$. For a local setting with pure seeds we have $\varsigma=\mathrm{diag}(\varsigma_a,\varsigma_b)$ with $|\varsigma_a|,|\varsigma_b|\le1$, and from Eq.~\eqref{eq:mobius} we obtain, with $\mathsf t=\tanh r$,
\begin{equation}
\Phi=\frac{1}{1-\mathsf t^2\varsigma_a\varsigma_b}\begin{pmatrix}(1-\mathsf t^2)\varsigma_a&\mathsf t(1-\varsigma_a\varsigma_b)\\\mathsf t(1-\varsigma_a\varsigma_b)&(1-\mathsf t^2)\varsigma_b\end{pmatrix}.
\label{eq:Phitmsv}
\end{equation}

\emph{Both arms.} For heterodyne detection of both modes we have $\Phi=\mathsf t\,\sigma_x$. As $\Sym^n(\sigma_x)$ exchanges the vectors with the labels $(m,n-m)$ and $(n-m,m)$, all singular values of $\Phi_n$ are equal to $\mathsf t^n$. The efficiencies of heterodyne detection of both modes in the order $n$ are then $(1\pm\tanh^nr)/2$, each $(n+1)$-fold, and Eq.~\eqref{eq:effpsi} becomes
\begin{equation}
\frac{F_Q}{F_{\mathrm{SLD}}}=\frac{F_W}{F_{\mathrm{SLD}}}=\frac12+\frac{\tanh^nr}{2\braket{\psi|\psi}}\,\mathrm{Re}\sum_{m=0}^n\overline{\psi_m}\,\overline{\psi_{n-m}} .
\label{eq:tmsveff}
\end{equation}
The equality with homodyne detection of both modes at uniform phases is the result of Sec.~\ref{sec:fisher-pure}, so this detection has the same efficiencies $(1\pm\tanh^nr)/2$, each $(n+1)$-fold. Note that entanglement plays no role in the spectrum~\eqref{eq:purespec}. Since heterodyne detection of every mode is invariant under passive linear optics and a balanced beam splitter turns $\ket{\Psi_r}$ into two single-mode squeezed vacua, the bound~\eqref{eq:purebound} is that of Eq.~\eqref{eq:sqhet} for $\bar n=0$. Local measurements, however, are not invariant under passive linear optics. For this reason the equality of local homodyne detection with heterodyne detection, as well as the two results on one arm and on local settings that follow, cannot be obtained from one mode in this way.

\emph{One arm.} If only the arm $a$ is measured, half of the efficiencies of every order vanish, and the other half are those of the thermal state of this arm. To show this, we note that for a measurement of the arm $a$ alone the setting does not obey Eq.~\eqref{eq:complete}, and we use $E_n=2^{n-1}\Pi_n\Sym^n(\Lambda)\Pi_n$ directly. For heterodyne detection we have $\Lambda=\tfrac12\big(\begin{smallmatrix}\openone&\mathsf t\sigma_z\\\mathsf t\sigma_z&\mathsf t^2\openone\end{smallmatrix}\big)$, for which $G_+\Lambda G_+/2$ acts on $\mathcal W_+$ as $\mathrm{diag}(1,\mathsf t^2)$ and $G_+\Lambda G_-/2$ maps $\mathcal W_-$ to $\mathcal W_+$ as $\mathsf t\,\sigma_x$. In this way we obtain
\begin{equation}
\frac{F}{F_{\mathrm{SLD}}}=\frac{1}{4\braket{\psi|\psi}}\sum_{m=0}^n\big|\sqrt{p_m}\,\psi_m+\sqrt{p_{n-m}}\,\overline{\psi_{n-m}}\big|^2
\label{eq:onearm}
\end{equation}
with the weights $p_m=q^{n-m}$ and $q=\mathsf t^2$. For homodyne detection of the arm $a$ at the phase $\theta$, $\Lambda$ projects onto a vector whose components along $\mathbf e_+^{(a)}$ and $\mathbf e_+^{(b)}$ are proportional to $e^{i\theta}\cosh r$ and $e^{-i\theta}\sinh r$. After averaging over $\theta$ we obtain Eq.~\eqref{eq:onearm} with $p_m=\mathbb P(X=n-m)$ for the binomial variable $X$ of Eq.~\eqref{eq:homkernel} with $\bar n=\sinh^2r$. In the form~\eqref{eq:onearm} only the components $m$ and $n-m$ are coupled, and the eigenvalues are $(p_m+p_{n-m})/2$ and $0$. Thus, if only one arm is measured, $n+1$ efficiencies of the order $n$ vanish. The other $n+1$ are the efficiencies~\eqref{eq:spectrum} of the thermal state of this arm, for heterodyne detection and for homodyne detection at a uniform phase, in the modes with $k_0=\min(m,n-m)$ and $|l|=|2m-n|$.

We now show that the visible directions are those of the arm. For a direction $\Delta_a$ of the thermal state $\rho_a$ of the arm $a$ with the SLD $L_a$, the direction $\Delta=[(L_a\otimes\openone)\rho_0+\rho_0(L_a\otimes\openone)]/2$ reduces to $\Delta_a$, and for this direction $F_{\mathrm{SLD}}(\Delta)=\tr(\rho_aL_a^2)=F_{\mathrm{SLD}}(\Delta_a)$. In the inner product of the SLDs (Appendix~\ref{app:structure}) this direction is orthogonal to every direction that does not change the state of the arm, so that it is measured in the arm $a$ with the efficiency of $\Delta_a$. In this way we connect the two-mode squeezed vacuum to Sec.~\ref{sec:fisher-thermal}. Since $\ket{\Psi_r}$ purifies the thermal state of one arm, half of the directions of each order do not change the state of this arm and are invisible in it, whereas the other half is measured, direction by direction, with the efficiencies of the thermal state.

\emph{Local settings.} General-dyne detection with the seed $\Sigma=V$ is Fisher-symmetric at $\ket{\Psi_r}$. Unfortunately, this seed is entangled, and we show that no measurement with local seeds, shared randomness and added noise can replace it. Let the settings be local, i.e., general-dyne or homodyne detection of each arm or no measurement of an arm, and let them be drawn at random independently of the outcomes and recorded. Such a measurement will be called nonadaptive. No such measurement, with or without Gaussian noise added to its outcomes, has the efficiency $1/2$ in all directions of the orders $1$ and $2$. We first assume that all seeds are pure. In this case, by Eq.~\eqref{eq:pair}, the efficiency $1/2$ in all directions of the orders $1$ and $2$ requires $\mathbb E\,\Phi=0$ and $\mathbb E\,\Sym^2(\Phi)=0$. The off-diagonal element of Eq.~\eqref{eq:Phitmsv} is $\mathsf t(1-\varsigma_{ab})/(1-\mathsf t^2\varsigma_{ab})$ with $\varsigma_{ab}=\varsigma_a\varsigma_b$, $|\varsigma_{ab}|\le1$, and
\begin{equation}
\mathrm{Re}\,\frac{1-\varsigma_{ab}}{1-\mathsf t^2\varsigma_{ab}}\ge\frac{(1-|\varsigma_{ab}|)(1-\mathsf t^2|\varsigma_{ab}|)}{|1-\mathsf t^2\varsigma_{ab}|^2}\ge0,
\label{eq:nolocal}
\end{equation}
where the left-hand side vanishes only for $\varsigma_{ab}=1$. Hence $\mathbb E\,\Phi=0$ requires $\varsigma_{ab}=1$ with probability one, i.e., homodyne detection of both arms with $\theta_a+\theta_b=0$ modulo $\pi$. Then $\Phi=\mathrm{diag}(\varsigma_a,\bar\varsigma_a)$, and for every setting the element of $\Sym^2(\Phi)$ between the vectors with the label $(1,1)$ equals $\varsigma_a\bar\varsigma_a=1$. Consequently $\mathbb E\,\Sym^2(\Phi)\ne0$. A local setting with a mixed seed, without a measurement of one arm, or with Gaussian noise added to the outcomes is obtained from a local setting with pure seeds by adding noise to its outcomes or by discarding them. The reason is that every covariance matrix of one mode that obeys the uncertainty relation exceeds that of a pure state by a nonnegative matrix. The Fisher information of such a setting is therefore not larger in any direction. If such a measurement had the efficiency $1/2$ in all directions of the orders $1$ and $2$, the settings with pure seeds would have at least $1/2$ there and, since their mean is $1/2$ by Eq.~\eqref{eq:pair}, exactly $1/2$, which we excluded. However, measurements whose settings depend on earlier outcomes of the other arm are not covered by this statement. Both orders are needed for this statement. Indeed, homodyne detection of both arms at the phases $\theta_a$ and $\theta_b=-\theta_a$ with a uniform $\theta_a$ has the efficiency $1/2$ in all directions of the first order.

\begin{figure}[t]
\centering\includegraphics[width=\columnwidth]{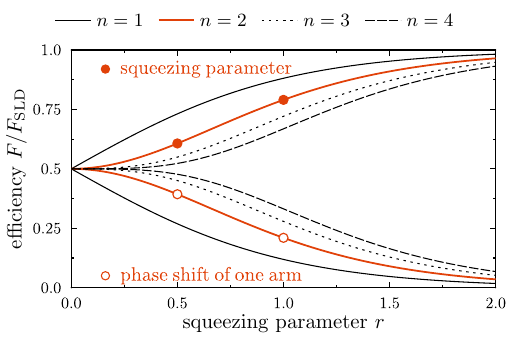}
\caption{Efficiencies of heterodyne detection of both modes and of homodyne detection of both modes at uniform phases near the two-mode squeezed vacuum [Eq.~\eqref{eq:tmsveff}]. Lines: the efficiencies $(1\pm\tanh^nr)/2$ of the orders $n=1,\dots,4$, between which the efficiency of every direction of the order $n$ lies. Symbols: a change of the squeezing parameter (filled) and a phase shift of one arm (open) at $r=0.5$ and $r=1$. Both directions belong to the second order.}\label{fig:tmsv}
\end{figure}

\emph{Physical directions.} We now use Eq.~\eqref{eq:tmsveff} to find the efficiencies of the two local detections for physical changes of $\ket{\Psi_r}$ (Fig.~\ref{fig:tmsv}). Since the generator of Eq.~\eqref{eq:tmsv} commutes with $U$, we have $\partial_r\ket{\Psi_r}=U\ket{1,1}$, i.e., a direction of the second order with $\psi_1=1$. The phase shift of one arm, $-i[\hat a^\dagger\hat a,\rho_0]$, has $\ket\psi=-i\cosh r\sinh r\,U\ket{1,1}$ and $F_{\mathrm{SLD}}=\sinh^22r$. A change of the squeezing parameter and a phase shift of one arm thus lie in the same mode of the second order, with opposite signs of $\psi_1^2$. Their efficiencies are $(1+\tanh^2r)/2$ and $(1-\tanh^2r)/2$, i.e., 0.607 and 0.393 at $r=0.5$ and 0.790 and 0.210 at $r=1$, so the phase shift is measured less efficiently as the squeezing grows. Since $\hat a\ket{\Psi_r}=\sinh r\,U\ket{0,1}$ and $\hat a^\dagger\ket{\Psi_r}=\cosh r\,U\ket{1,0}$, the admixture $\ket{\Psi_r}+\varepsilon\hat a\ket{\Psi_r}$ of the photon-subtracted state, and likewise of the photon-added state, is a direction of the first order with a single component. For such a direction the sum in Eq.~\eqref{eq:tmsveff} vanishes, and the efficiency is exactly $1/2$. For fixed settings the efficiency is distributed differently. In homodyne detection of two commuting quadratures at fixed phases, local or joint, the efficiency has only the stationary values $1$ and $0$ (Theorem~\ref{thm:single}), and a direction that is not an eigenvector lies between them. For example, the measurement of $\hat x_a-\hat x_b$ and $\hat p_a+\hat p_b$ behind a balanced beam splitter has the efficiency $1$ for the squeezing parameter and $0$ for the phase shift. The first value agrees with the Fisher information of this measurement found in Ref.~\cite{Bressanini2024}, which equals the quantum Fisher information of $r$.

Since loss in one arm admixes the photon-subtracted state incoherently, it is not a direction of finite $F_{\mathrm{SLD}}$. The efficiency of every Gaussian measurement is then zero, and only the two detections can be compared with each other. The direction of the loss is $\hat a\rho_0\hat a^\dagger-(\hat a^\dagger\hat a\rho_0+\rho_0\hat a^\dagger\hat a)/2$, and only the covariance matrix is changed by it. For this direction and for the fixed settings above we use the Fisher information $\tfrac12\tr[(\Xi\,\partial\Xi^{-1})^2]$ of a normal distribution. Here $\Xi^{-1}=HVH^{\mathsf T}+\Sigma$ is the covariance matrix, and $\partial\Xi^{-1}$ is its derivative with respect to the parameter. For the transmissivity $\eta$ of the arm $a$ we take $1-\eta$ as the parameter, and at $\eta=1$ we find
\begin{equation}
F_Q=\frac{\bar n}{2},\qquad F_W=\bar n\Big(1-\frac{1}{2\cosh2r}\Big),
\label{eq:tmsvloss}
\end{equation}
so homodyne detection at uniform phases carries more information than heterodyne detection, by a factor between $1$ and $2$. For homodyne detection of $\hat x_a$ and $\hat x_b$ at fixed phases we obtain $\bar n(3\bar n+1)$. This value agrees with the local homodyne detection of Ref.~\cite{Radia2026} without thermal background.

At a mixed reference state the two local detections differ already for displacements, e.g., for the two-mode squeezed thermal state whose covariance matrix is $2v$ times that of $\ket{\Psi_r}$. Heterodyne detection of both modes then obeys the bound~\eqref{eq:sqhet} with $2v$ in place of $2\bar n+1$. In contrast, for homodyne detection of both modes at uniform phases the first-order efficiencies remain $(1\pm\tanh|r|)/2$ for every $v$. This is because in the coordinates $\mathbf z=(\sqrt vS)^{-1}\mathbf x$ the matrix $\Lambda$ of a homodyne setting does not depend on $v$, and $J_1=\openone$. A bound on such homodyne detection in all orders is, however, beyond the scope of this paper, because Theorem~\ref{thm:mix} uses the angular momentum of one mode.

\subsection{Distant states and \texorpdfstring{$\chi^2$}{chi-squared} divergences}\label{sec:fisher-chi2}
The forms $F$ and $F_{\mathrm{SLD}}$ are quadratic in $\Delta$ and do not require $\Delta$ to be small. For any state $\rho$ with the outcome density $f_\rho$,
\begin{equation}
F(\rho-\rho_0)=\int\frac{f_\rho^2}{f_0}-1=\chi^2(f_\rho\|f_0),
\label{eq:chi2}
\end{equation}
the $\chi^2$ divergence of the outcome distributions, and $F_{\mathrm{SLD}}(\rho-\rho_0)$ is the quantum $\chi^2$ divergence that belongs to the SLD (Bures) metric~\citep{LesniewskiRuskai1999,Temme2010}. Whenever the latter is finite, Theorems~\ref{thm:structure}--\ref{thm:mix} and the results derived from them hold with $\Delta=\rho-\rho_0$ at any distance from the reference state. For example, near a thermal state of one mode, $\chi^2(f_\rho\|f_0)$ is at most $1/2$ times the quantum divergence for homodyne detection at a uniformly random phase, and at most $\lambda$ times this divergence for heterodyne detection. However, for a pure reference state the quantum divergence is infinite for every $\rho\ne\rho_0$, and this statement is empty.

\begin{figure}[t]
\centering\includegraphics[width=\columnwidth]{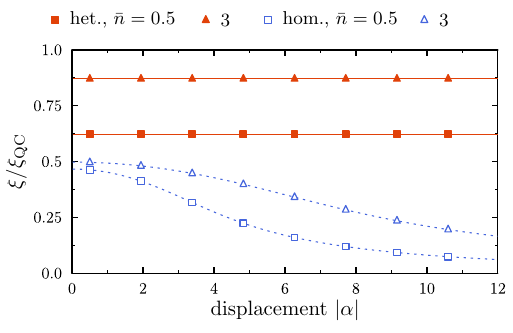}
\caption{Chernoff exponents for a displacement of a thermal state by the coherent amplitude $\alpha$, divided by the quantum Chernoff exponent $\xi_{\mathrm{QC}}$ of a collective measurement [Eq.~\eqref{eq:dispchernoff}]. Solid lines and filled symbols: heterodyne detection. Broken lines and open symbols: homodyne detection at a uniformly random recorded phase. Squares: $\bar n=0.5$, triangles: $\bar n=3$.}\label{fig:dispchernoff}
\end{figure}

For a displacement of a thermal state of one mode by a coherent amplitude $\alpha$ of any size, we can write the Chernoff information of the two quadrature detections itself in a closed form. For heterodyne detection ($\xi_Q$) and for homodyne detection at a uniformly random phase ($\xi_W$) we obtain
\begin{equation}
\xi_Q=\frac{|\alpha|^2}{4(\bar n+1)},\qquad\xi_W=u-\ln I_0(u),\quad u=\frac{|\alpha|^2}{4(2\bar n+1)}.
\label{eq:dispchernoff}
\end{equation}
Here $I_0$ is the modified Bessel function of the first kind of order zero. We compare them with the quantum Chernoff exponent of a collective measurement on many copies (defined in Sec.~\ref{sec:fisher-global})~\citep{Audenaert2007,Calsamiglia2008,PirandolaLloyd2008,Tan2008}. For the displacement this exponent equals $\xi_{\mathrm{QC}}=|\alpha|^2(\sqrt{\bar n+1}-\sqrt{\bar n})^2$. The largest exponent that Eq.~\eqref{eq:localchernoff} allows a measurement on single copies is $\varepsilon^2F_{\mathrm{SLD}}/8=|\alpha|^2/[2(2\bar n+1)]$. For $\bar n>0$ and small $|\alpha|$ the collective exponent exceeds it by the factor $2(2\bar n+1)(\sqrt{\bar n+1}-\sqrt{\bar n})^2$, which falls from $2$ for $\bar n\to0$ to $1$ for large $\bar n$. The ratios to $\xi_{\mathrm{QC}}$ in Fig.~\ref{fig:dispchernoff} are therefore smaller than the efficiencies of Sec.~\ref{sec:fisher-thermal}. For small $|\alpha|$ the ratio $\xi_W/\xi_Q$ is $(\bar n+1)/(2\bar n+1)$, the ratio of the displacement efficiencies $1/2$ and $\lambda$. For large $|\alpha|$ the heterodyne exponent keeps growing as $|\alpha|^2$, whereas $\xi_W\approx\tfrac12\ln(2\pi u)$ grows only logarithmically. This is because the error is then dominated by the events recorded at phases at which the displacement is not visible. The slow growth is a property of a phase that is drawn at random for every event. If a fixed share of the events is recorded at each of $K\ge2$ equally spaced phases, the average over the phases stands outside the logarithm (Sec.~\ref{sec:fisher-global}). In this case the exponent is $u$ for every $|\alpha|$, i.e., $(\bar n+1)/(2\bar n+1)$ times the heterodyne exponent. In Fig.~\ref{fig:dispchernoff} the ratio for heterodyne detection does not depend on $|\alpha|$, and the one for homodyne detection decreases as $|\alpha|$ grows. The exponents are derived in Appendix~\ref{app:chi2}, and we checked Eq.~\eqref{eq:dispchernoff} against direct integrals to a relative accuracy of $10^{-9}$.

\subsection{Noise}\label{sec:fisher-noise}
We now include loss and noise in front of the detector, which change both the reference state and the modes of a direction. Consider a Gaussian channel that maps the phase-space point $\mathbf x$ to $T\mathbf x$ plus independent Gaussian noise of covariance $N_c$. Under such a channel the covariance matrix $V$ of a Gaussian state becomes $V'=TVT^{\mathsf T}+N_c$. Using the same conditional expectation as in Theorem~\ref{thm:structure}, we find that the modes are mapped order by order,
\begin{equation}
g_n\to\mathrm{Sym}^n(C)\,g_n,\qquad C=V'^{-1/2}\,T\,V^{1/2},
\label{eq:channelmodes}
\end{equation}
where the new modes refer to the new state. Since the matrix $C$ does not increase the length of any vector, no mode is amplified by the channel. For photon loss with transmissivity $\eta$ we have $T=\sqrt\eta\,\openone$ and $N_c=(1-\eta)\openone/2$. Gaussian noise that displaces the state at random with variance $\sigma^2$ per quadrature is described by $T=\openone$ and $N_c=\sigma^2\openone$. We obtain the Fisher informations after the channel from Theorem~\ref{thm:structure} for the state with the covariance matrix $V'$. For a thermal state of one mode Eq.~\eqref{eq:channelmodes} reduces to
\begin{equation}
b_{n,l}\to
\begin{cases}
\big(\tfrac{v}{v+\sigma^2}\big)^{n/2}b_{n,l}, & \text{Gaussian noise},\ \bar n\to\bar n+\sigma^2,\\[2pt]
\big(\tfrac{\eta v}{\eta\bar n+1/2}\big)^{n/2}b_{n,l}, & \text{loss},\ \bar n\to\eta\bar n,\\[2pt]
e^{-\gamma l^2/2}\,b_{n,l}, & \text{dephasing},\ \bar n\to\bar n,
\end{cases}
\label{eq:noisemodes}
\end{equation}
where we added dephasing by a Gaussian random phase of variance $\gamma$. This dephasing is not a Gaussian channel. Nevertheless, a thermal state is left unchanged by it, and the angular component $l$ is multiplied by $e^{-\gamma l^2/2}$ (Appendix~\ref{app:noise}). We quote the strength $\nu$ of the three channels in percent. For loss it is $\nu=100(1-\eta)\%$, for dephasing we take $\gamma=3\nu/100$, and for Gaussian noise $\sigma^2=\nu/100$. We checked Eq.~\eqref{eq:noisemodes} against direct integrals. The relative accuracy was $10^{-13}$ for loss and dephasing and $10^{-10}$ for Gaussian noise.

Since the factors for loss and Gaussian noise are the same for both detections, these channels change the comparison~\eqref{eq:criterion} only through the occupation. Under loss it becomes the effective occupation $\bar n_{\rm eff}=\eta\bar n$, and under Gaussian noise $\bar n+\sigma^2$. With loss the comparison moves toward the vacuum, where homodyne detection is never worse. For Gaussian noise it is shifted toward large $\bar n$, and there more modes favor heterodyne detection. In the case of dephasing the reference state and the bounds~\eqref{eq:thermalbounds} remain unchanged. Every mode with $l\ne0$ is damped by dephasing, most strongly those with $|l|\ge2$, which include both the edge modes that favor heterodyne detection and the modes of high order with $2\le|l|\ll n$ that favor homodyne detection. As a result, the net effect of dephasing depends on which modes carry the difference between the states.
\section{Test on distant states}\label{sec:test}
The relation~\eqref{eq:localchernoff} between the Fisher information and the error exponent holds for small changes, whereas the states that a check of a non-Gaussian source must tell apart are far from each other. We therefore test how far the statements of the theory carry on a task in which the number of events is limited. The task is the identification of the family of a state from a record of $N$ independent detection events, each from one preparation of the state. Such a decision is made when one checks that a heralded source of GKP states~\citep{Konno2024} still prepares the intended state. In heterodyne detection each event is a point $(x,p)$. In homodyne detection it is one quadrature value at one of $K=36$ equally spaced phases, and the $N$ events are divided among these phases as evenly as possible. The information contained in the records is given by the Chernoff exponents (Sec.~\ref{sec:fisher-global}). We then use a likelihood-ratio observer, i.e., the decision that uses all of this information (Sec.~\ref{sec:res-lr}), and in Sec.~\ref{sec:test-networks} we show with neural networks what a trained classifier achieves in practice.

The base set consists of seven families of states of one mode, Fock, coherent, squeezed vacuum, thermal, cat, GKP and binomial states~\citep{GKP2001,Michael2016}. All families have the same mean photon numbers, $2$, $3$, $4$ and $5$, which occur with the same frequencies in every family. In this way the mean photon number carries no information about the family. The phases and the squeezing axes are random. A second set of seven families, with numerically optimized code states instead of the squeezed vacuum and with unequal mean photon numbers, follows Ahmed et al.~\cite{Ahmed2021}. The states pass through loss, dephasing or Gaussian noise (Sec.~\ref{sec:fisher-noise}), and we evaluate both detections on identical states. For heterodyne detection the outcomes are binned into $128\times128$ pixels on $[-12,12]^2$, and for homodyne detection into $128$ bins at each phase. The samplers of the states and all parameters are given in the repository~\cite{QMIcode}.

\subsection{Chernoff exponents}\label{sec:fisher-global}
We calculate the Chernoff information~\eqref{eq:chernoff} of both detections on the pixel grid. For each family we took 24 test states, and the information has been calculated between each of them and the closest state of the other families (24 states per family). For homodyne detection with $N/K$ events at each phase the exponent is the maximum over $s$ of the average over the phases of $-\ln\int f_0^{\,s}f_1^{\,1-s}$. Here $f_0$ and $f_1$ are the outcome densities of the two states at one phase, and we denote this exponent again by $\xi_W$. It differs from the exponent for a phase drawn at random for every event, Eq.~\eqref{eq:dispchernoff}, in which the average over the phases stands inside the logarithm. For small changes of the state the two exponents agree. For every pair $\rho_i,\rho_j$ we also calculate the quantum Chernoff exponent of a collective measurement~\citep{Audenaert2007,Nussbaum2009,Calsamiglia2008},
\begin{equation}
\xi_{\mathrm{QC}}=-\ln\min_{0\le s\le1}\tr\big(\rho_i^{\,s}\rho_j^{\,1-s}\big).
\label{eq:qchernoff}
\end{equation}
On noiseless states the median of the ratio $\xi_W/\xi_Q$ (range over the two sets) is 1.12\mbox{--}1.21. Under $20\%$ dephasing ($\gamma=0.6$, a root-mean-square phase noise of $0.77$~rad) the median is 1.03\mbox{--}1.11. In the base set it takes the values $1.01$ and $1.25$ under $50\%$ and $80\%$ loss, and in the second set the values $0.87$ and $0.93$. The median ratio thus exceeds its value without noise only in the base set at $80\%$ loss. These medians do not show the shift toward homodyne detection that Sec.~\ref{sec:fisher-noise} gives for a fixed pair under loss. For the median pair the numbers of events needed by the two detections differ by at most 25\%. For both detections the median exponent lies in the range 8--23\% of the quantum exponent $\xi_{\mathrm{QC}}$ (over the two sets and these noise conditions), i.e., both detections stay far below the quantum limit, as the local efficiencies suggest.

\begin{figure}[t]
\centering\includegraphics[width=\columnwidth]{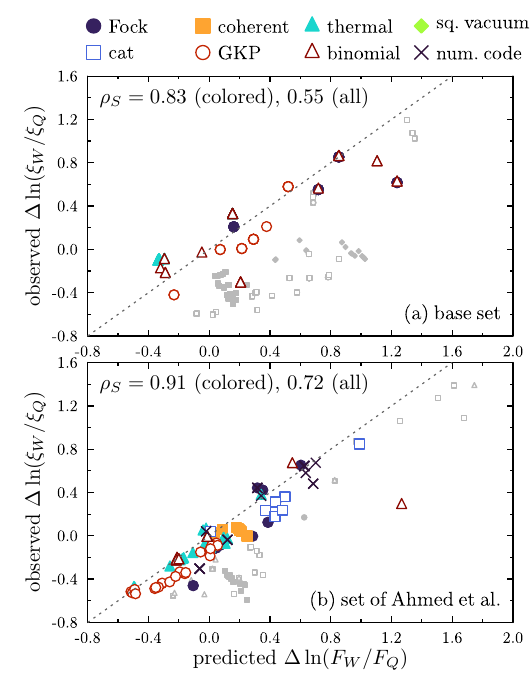}
\caption{Effect of $20\%$ dephasing on the ratio of the Chernoff exponents of homodyne and heterodyne detection for fixed pairs of states. Each point is one test state $\rho_i$ and its closest state $\rho_j$ of another family without noise. Vertical axis: observed change of $\ln(\xi_W/\xi_Q)$ under dephasing (here $\Delta$ denotes a difference). Horizontal axis: change of $\ln(F_W/F_Q)$ predicted by Eqs.~\eqref{eq:modes} and~\eqref{eq:noisemodes} for the direction $\rho_j-\rho_i$ at the thermal state with the mean photon number of the pair. (a) Base set, (b) set of Ahmed et al.~\cite{Ahmed2021}. Dotted line: equality. Large colored symbols: pairs within the range of validity of the theory, for which dephasing changes the mean heterodyne distribution of the pair by at most $0.15$ in total-variation distance. Small gray symbols: the other pairs. Each panel gives the Spearman correlation coefficient, denoted $\rho_S$ in the panels, for the pairs within the range (colored) and for all pairs. Symbols mark the family of $\rho_i$ (num.\ code: the numerically optimized code states of the second set).}\label{fig:dephasing}
\end{figure}

In the base set the median ratio decreases from 1.21 to 1.03 upon dephasing. A part of this decrease is a change of the task. The reason is that for 65\% of the test states the closest family under heterodyne detection is, after dephasing, a different one. We have also fixed the pairs, taking a state and its closest state under heterodyne detection without noise. For these fixed pairs the median ratio changes only from 1.55 to 1.47 (the value without noise exceeds 1.21 because in the first comparison each detection has its own closest state). For each of these pairs we can compare the observed change with the prediction of the theory of Sec.~\ref{sec:fisher}. To this end we take the difference $\rho_j-\rho_i$ as a direction at the thermal state $\rho_0$ with the mean photon number of the pair. By Eqs.~\eqref{eq:modes} and~\eqref{eq:noisemodes}, dephasing leaves $\rho_0$ unchanged and multiplies the modes of angular momentum $l$ by $e^{-\gamma l^2/2}$. From this we obtain the predicted change of $\ln(F_W/F_Q)$ for the pair (Fig.~\ref{fig:dephasing}). Without noise, the pairs are ranked by the ratio $F_W/F_Q$ almost as by the ratio $\xi_W/\xi_Q$. In the two sets the Spearman rank correlation coefficients are $0.88$ and $0.94$. The predicted change under dephasing is correlated with the observed one (Spearman correlations of $0.55$ and $0.72$). However, many pairs lie below the diagonal. In the base set the median predicted change is $+0.26$, against the observed $-0.06$.

The offset between the predicted and the observed change comes from the reference state. The prediction uses a thermal state, which dephasing does not change, whereas the states of a pair are dephased themselves. If we evaluate the same quadratic forms with the mean outcome distribution of the pair in place of that of $\rho_0$, their change reproduces the observed one. The Spearman correlations are then $0.98$ and $0.96$, and in the base set the median change is $-0.03$. The offset marks the range of validity of the theory. As a measure of this range we take the total-variation distance (half of the summed absolute difference of two distributions on the pixel grid) by which dephasing changes the mean heterodyne distribution of a pair. This distance is at most $0.15$ for $93$ and $126$ of the $168$ pairs of each of the two sets. For these pairs the prediction follows the observed change, with Spearman correlations of $0.83$ and $0.91$. Their median offsets are $+0.01$ and $-0.09$, while for the other pairs we obtained median offsets of $-0.63$ and $-0.47$. In the base set the pairs within the range are those of the Fock, thermal and binomial states and most of those of the GKP states, whereas all pairs of the coherent, squeezed vacuum and cat states lie outside it.

\subsection{Optimal decision on the events}\label{sec:res-lr}
Once the phases and axes of the coherent, squeezed and cat states are put on a grid of $2.5^\circ$, the base set contains finitely many prototype states per family. The probability of every record is therefore known for each channel and strength. We use a likelihood-ratio observer that knows the prototypes and the channel. For a given record the observer computes the multinomial likelihood under each prototype and sums it inside each family with the sampling weights of the base set. It then chooses the family with the largest sum. In our simulations $5000$ records have been generated per family, channel and $N$. We give $95\%$ percentile bootstrap intervals over test states, stratified by family, with $4000$ resamples. The error of this observer is an upper bound on the smallest error that any classifier can reach for these known states. For the binned records the two errors differ only through the grid of $2.5^\circ$.

\begin{figure}[t]
\centering\includegraphics[width=\columnwidth]{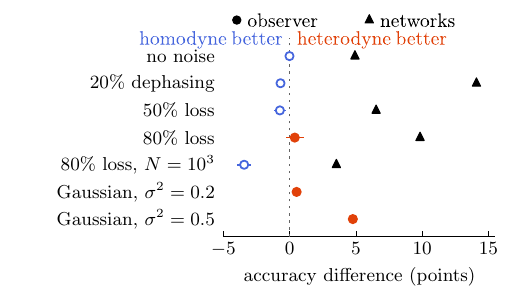}
\caption{Difference of the accuracies of the likelihood-ratio observer on heterodyne and on homodyne records of the base set, in percentage points, for records of $N=100$ events and, for $80\%$ loss, also of $N=10^3$ events. Circles: the observer, filled if heterodyne detection is better and open if homodyne detection is better, with $95\%$ bootstrap intervals over test states (bars). Triangles: the same difference for the neural networks on heterodyne histograms and on reconstructed images. The networks show what a trained classifier achieves in practice (Sec.~\ref{sec:test-networks}), and they were not evaluated in the last two rows. The key above the panel names the markers.}\label{fig:observer}
\end{figure}

At $N=100$ the observer shows no advantage of heterodyne detection under loss and dephasing (Fig.~\ref{fig:observer}). Under $20\%$ dephasing its errors are $4.5\%$ on heterodyne records and $3.8\%$ on homodyne records. Under $50\%$ loss these errors are $8.2\%$ and $7.4\%$, and under $80\%$ loss they reach $33.1\%$ and $33.4\%$. Positive values of the difference of the accuracies on heterodyne and homodyne records favor heterodyne detection. Without noise this difference is $-0.02$ percentage points ($[-0.04,-0.01]$). Under $20\%$ dephasing we obtained $-0.7$ points ($[-1.0,-0.4]$), and under $50\%$ loss again $-0.7$ points ($[-1.2,-0.3]$). At $80\%$ loss the difference is $+0.4$ points ($[-0.3,1.0]$). For the two channels that shift the effective occupation the observer agrees with the theory in the sign of its preference. At $80\%$ loss and $N=10^3$ it prefers homodyne detection by $3.4$ points. This is expected when loss moves the comparison toward the vacuum, where homodyne detection is never worse. Under Gaussian noise with $\sigma^2=0.5$, i.e., one unit of vacuum noise in each quadrature, the observer prefers heterodyne detection at $N=100$ by $4.8$ points ($[4.4,5.1]$). At $\sigma^2=0.2$ this preference is $0.5$ points ($[0.4,0.7]$). Both preferences are expected if the noise raises the effective occupation. Since the error changes by at most about $0.5$ points between $K=36$, $12$ and $6$ phases, the number of phases does not limit the information in the records for these families.

\subsection{Classifiers in practice}\label{sec:test-networks}
The observer uses all the information in the records. In practice, however, the states are not known and a classifier must learn them from examples. Such a classifier need not use all the information, and a comparison of two detections by trained classifiers can then differ from the comparison of the detections themselves. We demonstrate this relation of practice to theory with neural networks. Neural networks have been used to reconstruct quantum states~\citep{Torlai2018,Carrasquilla2019}, also from homodyne records~\citep{Tiunov2020}, and to classify optical states from images of their phase-space functions~\citep{Ahmed2021}. They have also been used to detect nonclassicality and Wigner negativity from histograms of homodyne events without a reconstruction~\citep{Cimini2020,Gebhart2021} (for a review see Ref.~\cite{Gebhart2023}). We use convolutional neural networks~\citep{LeCun1998} of $2.2\times10^6$ weights. They have been trained on records of the same task, in three to five independent training runs for each input. As the input we take the histogram of the heterodyne events, the image of the Wigner function reconstructed from the homodyne events by filtered back-projection~\citep{LvovskyRaymer2009}, or the histograms of the homodyne events without a reconstruction. All networks are tested on the same test states, and for the difference of the accuracies of two networks we give $95\%$ intervals that combine the variation over the test states and over the training runs.

The networks do not reach the optimal decision. At $N=100$ and without noise the network on heterodyne histograms has an error of $4.1\%$, whereas the observer errs in $0.03\%$ of the cases. Under $20\%$ dephasing the error of this network is $17.3\%$, against $4.5\%$ of the observer. Under $50\%$ loss we obtained $18.7\%$ and $8.2\%$, and under $80\%$ loss $41.7\%$ and $33.1\%$.

How much of the information a network uses depends on its input. In the base set at $N=100$ the network on reconstructed images is less accurate than the network on heterodyne histograms. Without noise the difference is $4.9$ percentage points ($\ensuremath{[2.1,\,7.7]}$). Under $20\%$ dephasing it is $14.1$ points ($\ensuremath{[9.4,\,18.7]}$), and under $50\%$ and $80\%$ loss we obtained $6.5$ points ($\ensuremath{[3.0,\,10.0]}$) and $9.8$ points ($\ensuremath{[6.1,\,13.6]}$). For the same noise conditions the observer shows differences below one point (Fig.~\ref{fig:observer}). We have applied the Holm correction for multiple comparisons~\citep{Holm1979} over the $24$ compared cases (two sets, three values of $N$, four noise conditions). After this correction the differences under dephasing and loss remain significant at the level $0.05$ (adjusted $p=0.003$, $0.021$ and $0.004$), whereas the difference without noise is not significant ($p=0.054$). The gap disappears when the network is given the histograms of the homodyne events. In the four noise conditions the accuracy of this network differs from that of the network on heterodyne histograms by $0.0$, $\ensuremath{-}0.2$, $0.0$ and $2.3$ points, and all the $95\%$ intervals contain zero. In the second set no difference is significant after the correction. For $10^3$ events or more none is significant after the correction in either set. Since the observer shows no heterodyne advantage for these noise conditions and processing cannot add information, the heterodyne advantage of the networks at $N=100$ in the base set comes mostly from the classification of a reconstructed image and not from the detection.

The number of phases affects the network on reconstructed images but hardly the observer (Sec.~\ref{sec:res-lr}). This is expected from Sec.~\ref{sec:fisher-bounds} for directions whose modes have $|l|<K$ near a thermal state. Without noise and at $N=100$ the accuracy of this network rises from 91.0\% at $K=36$ to 93.0\% at $K=12$ and 94.3\% at $K=6$. The reason is that with fewer phases the histogram of each phase contains more events.

The difference between a network and the observer is thus the cost of learning and of the processing of the records, and not a property of the detection. For this reason we base the comparison of the two detections on the Fisher information, the Chernoff exponents and the observer, while trained classifiers do not enter it. A classifier used in practice can, in turn, be judged against the reference given by the theory and the observer.
\section{Discussion}\label{sec:discussion}
We now discuss what the theory implies for the choice between the two detections, how it relates to the known comparisons, and where it ceases to apply.

\emph{Choice of the measurement.} At the level of information we obtain from the theory a rule. We write the difference between two states in Hermite--Gauss modes around a thermal state with the occupation $\bar n_{\rm eff}$ that includes the noise. Under loss this occupation is $\eta\bar n$, and under Gaussian noise it is $\bar n+\sigma^2$. Using the criterion~\eqref{eq:criterion} with this occupation we can then tell mode by mode which detection carries more information. Dephasing does not change this rule. However, it damps the modes with $|l|\ge2$ most strongly, and in this way it changes which of them matter. In the base set the likelihood-ratio observer agrees with the rule in the sign of its preference under loss and Gaussian noise (Sec.~\ref{sec:res-lr}).

\emph{Relation to earlier work.} With the criterion~\eqref{eq:criterion} we place the known comparisons in one picture (Sec.~\ref{sec:fisher-thermal}). In finite dimensions Fisher-symmetric measurements were known~\citep{LiFerrie2016,ZhuHayashi2018}, and we have shown in Sec.~\ref{sec:fisher-squeezed} that both quadrature detections have this property at coherent states. The vanishing efficiencies of all modes with $k_0\ge1$ radial nodes as $\bar n\to0$ in Eq.~\eqref{eq:spectrum} agree, for changes of the covariance matrix, with the suboptimal scaling of Gaussian measurements for weak thermal sources~\citep{WangZhou2025}. To our knowledge, no earlier work gives the fraction of the quantum Fisher information that the two quadrature detections recover for arbitrary directions. We also know of no earlier bound on this fraction by a constant smaller than one for heterodyne detection and for homodyne detection at a uniformly random phase.

\emph{Limitations.} The theorems concern the Fisher information around Gaussian states and the $\chi^2$ divergences from them, and for the error exponents of distant states we rely on numerical Chernoff exponents. Between distant states without noise the theory of small changes ranks the pairs almost as the Chernoff exponents do. Its prediction of the effect of dephasing, however, holds only for pairs whose mean outcome distribution dephasing leaves almost unchanged, and under $20\%$ dephasing this condition holds for $93$ and $126$ of the $168$ pairs of each of the two sets (Sec.~\ref{sec:fisher-global}). The Chernoff exponents describe long records, whereas for short records the factor in front of the exponential matters, e.g., at $80\%$ loss the observer has no significant preference at $N=100$ and prefers homodyne detection at $N=10^3$. The theory gives the information in the records and not the accuracy of a classifier that must learn the states (Sec.~\ref{sec:test-networks}). The efficiency compares every direction with its own best measurement, and no single measurement attains the quantum Fisher information in all directions. For mixed reference states we give no bound on the average efficiency such as that of Sec.~\ref{sec:fisher-squeezed} for coherent states. The closed forms mostly concern one mode. For several modes we have Theorems~\ref{thm:structure} and~\ref{thm:single} and the pure states of Secs.~\ref{sec:fisher-pure} and~\ref{sec:fisher-tmsv}, but no counterpart of Theorem~\ref{thm:mix}. Bounds on local homodyne detection near mixed states of several modes and measurements with feed-forward between the modes are, however, beyond the scope of this paper, as are non-Gaussian measurements such as the displaced parity~\citep{Lutterbach1997}. The closed-form bounds assume an ideal detector, but a detector inefficiency can be included in the loss channel (the efficiencies then refer to the quantum Fisher information of the state after this loss) or in the added noise $\Sigma$ of Theorems~\ref{thm:structure} and~\ref{thm:single}. Unfortunately, neither the theory nor the test includes a mode mismatch or a drift of the phase of the local oscillator. The test of Sec.~\ref{sec:test} uses simulated records of a closed set of families.

\section{Conclusions}\label{sec:conclusions}
We have determined how much information homodyne and heterodyne detection recover about a small change of a Gaussian state, for every change and not only for the parameters of particular tasks. We have shown that near any Gaussian state of one or several modes neither the quantum Fisher information nor the Fisher information of a Gaussian measurement couples Hermite--Gauss modes of different order. Therefore, no change of the state is measured by a fixed Gaussian setting more efficiently than by the best displacement. Near a thermal state of one mode homodyne detection at a uniformly random phase recovers at most one half of the quantum Fisher information, and for heterodyne detection this fraction is at most $(2\bar n+1)/(2\bar n+2)$. Near a coherent state both detections are Fisher-symmetric, and squeezing of the reference state raises both bounds. At the two-mode squeezed vacuum, homodyne detection of both modes at uniformly random phases recovers the same information as heterodyne detection in every direction of finite quantum Fisher information, and no nonadaptive Gaussian measurement of each mode separately is Fisher-symmetric. In the base set of distant states the likelihood-ratio observer shows no heterodyne advantage under loss and dephasing at $N=100$ and a heterodyne advantage under Gaussian noise. Neural networks trained on reconstructed images show in the base set a heterodyne advantage under loss and dephasing that the records do not contain. We have used them to demonstrate how a classifier used in practice relates to the theory, and none of our conclusions about the two detections rests on them.

First, near a thermal state of one mode the criterion~\eqref{eq:criterion} puts the comparisons made task by task in one picture. For a given mode of the change and a given occupation it tells which detection carries more information. It reconciles the optimality of heterodyne detection for the joint estimation of a displacement with the advantage of homodyne detection in tomography. The bounds are, in addition, limits of Gaussian measurements on single copies. Near a thermal or a coherent state, a check of a source with homodyne detection at a uniformly random phase needs at least twice as many events as the measurement that is optimal for a known change. At a coherent state no measurement does better than this factor on average over the changes. Near a squeezed thermal state such a check needs at least $2/(1+\tanh|r|)$ times as many. For random non-Gaussian changes of up to seven photons near a thermal state both quadrature detections have median efficiencies 0.04\mbox{--}0.19 (Sec.~\ref{sec:fisher-thermal}) and need about $5$ to $25$ times as many. This factor is the gain available over the two standard detections to a measurement designed for the change, and a Gaussian setting tuned to the change can recover a part of it. Theorem~\ref{thm:structure}, finally, reduces the Fisher information of any Gaussian measurement near any Gaussian state to matrices of the size of phase space, and it can therefore be evaluated for other states of several modes and other Gaussian measurements.
\begin{acknowledgments}
We thank the members of the student team of the IDUB project who developed the qsdl library~\cite{qsdl}, which we used to generate the states of Sec.~\ref{sec:test}. This work was supported by the Excellence Initiative -- Research University (IDUB) program of Adam Mickiewicz University (Advanced Best Student Grant, Edition 5). K.B. acknowledges support from the EuroHPC Joint Undertaking under Horizon Europe Grant No.~101194322 (QEC4QEA), co-funded by the Polish National Centre for Research and Development (NCBiR) under Decision No.~DWM/EuroHPC/2023/429/2025.

We used the AI coding assistant Claude Code (Anthropic; version 2.1.308, with Claude Opus models, Claude Opus 5.5 in the final stage) to write and run the analysis code and the plotting scripts, to check the numbers in the text against the result files, and to draft and edit parts of the text. All theorems were proposed by K.B., and the AI assistant helped with their proofs, which we verified by reading them line by line and by numerical checks with code independent of the formulas. All numbers and figures are produced by our code from the result files, and no image in the paper is generated by an AI model. The authors reviewed the code, reran the analyses and checked every number and statement.
\end{acknowledgments}

\noindent\textit{Author contributions.} F.G., S.D.: software, investigation, data curation, visualization, writing (review and editing). K.B.: conceptualization, methodology, formal analysis, theorems and proofs, supervision, funding acquisition, writing (original draft, review and editing).

\noindent\textit{Competing interests.} The authors declare no competing interests.

\section*{Data availability}
All data are simulated and are regenerated from fixed random seeds by our code. The code, the scripts that generate every number and figure, and the result files behind them are in a public repository under the MIT license~\cite{QMIcode}. The networks of Sec.~\ref{sec:test-networks} are retrained by the same code, and the data sets (about $15$~GB) are not stored. The states of the base set of Sec.~\ref{sec:test} are generated with the qsdl library~\cite{qsdl} (commit a11056285ad9), which is not included in the repository. For the states of the second set we use our implementation of the samplers of the released code of Ahmed et al.~\cite{Ahmed2021code}.
\appendix
\section{Proofs and numerical checks}\label{app:proofs}
We prove here the three theorems of Sec.~\ref{sec:fisher}. The proofs rest on one analogy, which we set up in Appendix~\ref{app:conv}: the functions on phase space, weighted with the Wigner function of the reference state, behave as the states of $2M$ fictitious bosonic modes. It is the analogy with photons of Sec.~\ref{sec:fisher-modes}. To distinguish these bosons from the photons of the field, we speak of quanta. In this language the Hermite order $n$ is the sector of $n$ quanta. A Gaussian measurement, as well as the multiplication by a Gaussian state, acts on these quanta in the same way as a linear optical element acts on photons. Theorem~\ref{thm:structure} is proved in Appendix~\ref{app:structure} and the bounds in Appendix~\ref{app:bounds}. Some steps only verify that sums and integrals converge or that limits can be exchanged. Such steps are needed for completeness, but they can be skipped at first reading, and we mark them as technical points. In Appendix~\ref{app:local} we derive Eq.~\eqref{eq:localchernoff}. The numerical checks are collected in Appendix~\ref{app:checks}, while the formulas of Secs.~\ref{sec:fisher-chi2} and~\ref{sec:fisher-noise} are derived in Appendices~\ref{app:chi2} and~\ref{app:noise}.

\subsection{Local exponent}\label{app:local}
We first show that the Chernoff exponent of a small change of the state is the Fisher information. Let $f$ be the outcome density for $\rho_0$ and $f+\varepsilon\delta$ that for $\rho_0+\varepsilon\Delta$. A term of the second order in the state adds $\varepsilon^2\delta_2$ with $\int\delta_2=0$ to the density. Since $f$ vanishes nowhere for the Gaussian measurements of Sec.~\ref{sec:fisher-setting}, this term contributes $(1-s)\varepsilon^2\int\delta_2=0$ to the integral below and thus changes only the remainder. We expand $(f+\varepsilon\delta)^{1-s}=f^{1-s}(1+\varepsilon\delta/f)^{1-s}$ to the second order in $\varepsilon$. The term of the first order is proportional to $\int\delta$, and this integral vanishes because both densities are normalized. The term of the second order gives
\begin{equation}
-\ln\int f^s(f+\varepsilon\delta)^{1-s}=\frac{s(1-s)}{2}\,\varepsilon^2F+O(\varepsilon^3),
\label{eq:chernoffexp}
\end{equation}
with $F=\int\delta^2/f$. The factor $s(1-s)$ is largest at $s=1/2$, so the Chernoff exponent of Eq.~\eqref{eq:chernoff} is $\varepsilon^2F/8$. The expansion requires that $\int\delta^2/f$ is finite and, for the remainder, that $\int|\delta|^3/f^2$ is finite. Both conditions hold for $\Delta$ of finite Fock support near a thermal state and for every direction with a polynomial SLD, for which $\delta/f$ is a polynomial in the outcome. If the setting is drawn at random and recorded, the outcome is the pair of the setting and the value, and $F$ is then the average of the Fisher informations of the settings.

\subsection{Conventions and modes}\label{app:conv}
In this appendix we collect the conventions of phase space and of the reference state, and then we set up the analogy on which the proofs rest. In this analogy the Hermite orders are the sectors of $n$ fictitious quanta. With the rule of the coherent states we turn a calculation with exponential functions into the symmetric powers of Theorem~\ref{thm:structure}. In the last two paragraphs we give the bases in which $J_n$ is diagonal, for $M$ modes and for one mode.

\emph{Phase space.} We use $\hat{\mathbf x}=(\hat x_1,\hat p_1,\dots,\hat x_M,\hat p_M)$ with $[\hat x_j,\hat p_j]=i$ and the displacement (Weyl) operators $D_{\mathbf a}=e^{i\mathbf a\cdot\hat{\mathbf x}}$ with $\mathbf a\in\mathbb R^{2M}$. By the Baker--Campbell--Hausdorff formula they obey the composition law $D_{\mathbf a}D_{\mathbf b}=e^{-i\mathbf a^{\mathsf T}\Omega\mathbf b/2}D_{\mathbf a+\mathbf b}$, in which the phase factor is the only trace of the noncommutativity of the quadratures. The function $\tr(AD_{\mathbf b})$ of $\mathbf b$ is the characteristic function of an operator $A$ of finite trace norm (defined below), and the Wigner function is its Fourier transform, $W_A(\mathbf x)=(2\pi)^{-2M}\int\tr(AD_{\mathbf b})e^{-i\mathbf b\cdot\mathbf x}d^{2M}b$. Then $\tr A=\int W_A$ and $\tr(A^\dagger B)=(2\pi)^M\int\overline{W_A}W_B$. Different operators have different Wigner functions, and $|W_A|\le\|A\|_1/\pi^M$, where $\|A\|_1$ is the trace norm (the sum of the singular values of $A$). In the opposite direction, to a polynomial $h(\mathbf x)$ we assign the symmetrically ordered operator $\mathrm{Op}(h)$, defined by $\mathrm{Op}[(\mathbf a\cdot\mathbf x)^n]=(\mathbf a\cdot\hat{\mathbf x})^n$, for example $\mathrm{Op}(xp)=(\hat x\hat p+\hat p\hat x)/2$. We call such operators polynomial operators and $h$ the phase-space function (Weyl symbol) of $\mathrm{Op}(h)$. The phase-space function of $D_{\mathbf a}$ is $e^{i\mathbf a\cdot\mathbf x}$.

\emph{Reference state.} The reference state $\rho_0$ has $W_0=\mathcal N(0,V)$, the normal density with zero mean and the covariance matrix $V$, and the characteristic function $\tr(\rho_0D_{\mathbf a})=e^{-\mathbf a^{\mathsf T}V\mathbf a/2}$. By the Williamson theorem (the normal-mode decomposition; see, e.g., Ref.~\cite{Radia2026}) there is a symplectic matrix $S$, $S\Omega S^{\mathsf T}=\Omega$, with $V=SV_0S^{\mathsf T}$ and $V_0=\mathrm{diag}(v_1,v_1,\dots,v_M,v_M)$. Physically, a Gaussian unitary operation (a network of beam splitters, phase shifters and squeezers) transforms $\rho_0$ into a product of the thermal states $(1-q_j)q_j^{\nhat_j}$ of $M$ modes with the photon-number operators $\nhat_j$, which we call the normal modes. We write $\ket{\Psi_{\mathbf m}}$ for the eigenvectors of $\rho_0$, the images of the Fock states of the normal modes, and $\pi_{\mathbf m}=\prod_j(1-q_j)q_j^{m_j}$ for its eigenvalues, with $\mathbf m=(m_1,\dots,m_M)$ and $|\mathbf m|=\sum_jm_j$. Since we take $0^0=1$, we have $\pi_{\mathbf m}=0$ unless $m_j=0$ in every pure normal mode. We set $\tau_j=1/(2v_j)\in(0,1]$ and $q_j=(1-\tau_j)/(1+\tau_j)$. For one mode we omit the index $j$, write $m$ for $\mathbf m$, and have $V=vSS^{\mathsf T}$, $\pi_m=(1-q)q^m$ and $\kappa=1-\tau$.

\emph{Hermite orders as sectors of quanta.} We consider the complex functions on phase space with the inner product $\langle g,h\rangle=\int\bar gh\,W_0$ and finite norm. For them we use the normalized quadratures $\mathbf z=V^{-1/2}\mathbf x$ of Sec.~\ref{sec:fisher-modes}, which under $W_0$ are uncorrelated and have zero mean and unit variance. For a complex vector $\boldsymbol\gamma\in\mathbb C^{2M}$ we define the function
\begin{equation}
e_{\boldsymbol\gamma}(\mathbf x)=\exp\big(\boldsymbol\gamma\cdot\mathbf z-\tfrac12\boldsymbol\gamma\cdot\boldsymbol\gamma\big),\qquad e_{t\boldsymbol\gamma}=\sum_{n\ge0}\frac{t^n}{n!}\wick{(\boldsymbol\gamma\cdot\mathbf z)^n},
\label{eq:genf}
\end{equation}
where the dot is the product without complex conjugation, $\boldsymbol\gamma\cdot\boldsymbol\delta=\sum_i\gamma_i\delta_i$ for two vectors with the components $\gamma_i$ and $\delta_i$. The second relation is the expansion of $e_{t\boldsymbol\gamma}$ in powers of the number $t$. It defines the polynomials $\wick{(\boldsymbol\gamma\cdot\mathbf z)^n}$ of degree $n$ in $\mathbf z$, which are called Wick powers. For one variable Eq.~\eqref{eq:genf} is the generating function of the Hermite polynomials, and for a real unit vector $\boldsymbol\gamma$ the Wick power is the Hermite polynomial of degree $n$ of the quadrature $\boldsymbol\gamma\cdot\mathbf z$, e.g., $\wick{z_1^2}=z_1^2-1$. For two numbers $s$ and $t$ we obtain from a Gaussian integral $\langle e_{s\boldsymbol\gamma},e_{t\boldsymbol\delta}\rangle=e^{\bar st\,\bar{\boldsymbol\gamma}\cdot\boldsymbol\delta}$. By comparing the coefficients of $\bar s^mt^n$ on both sides we find
\begin{equation}
\big\langle\wick{(\boldsymbol\gamma\cdot\mathbf z)^m},\wick{(\boldsymbol\delta\cdot\mathbf z)^n}\big\rangle=\delta_{mn}\,n!\,(\bar{\boldsymbol\gamma}\cdot\boldsymbol\delta)^n .
\label{eq:wickgram}
\end{equation}
Here $\delta_{mn}$ is the Kronecker symbol. The same relations are obeyed by $2M$ bosonic modes with the annihilation operators $\hat{\mathfrak b}_1,\dots,\hat{\mathfrak b}_{2M}$ and the vacuum $\ket{\mathrm{vac}}$. For these modes the overlap of the unnormalized coherent states $e^{\boldsymbol\gamma\cdot\hat{\boldsymbol{\mathfrak b}}^\dagger}\ket{\mathrm{vac}}$ and $e^{\boldsymbol\delta\cdot\hat{\boldsymbol{\mathfrak b}}^\dagger}\ket{\mathrm{vac}}$ is $e^{\bar{\boldsymbol\gamma}\cdot\boldsymbol\delta}$. The overlap of the states $(\boldsymbol\gamma\cdot\hat{\boldsymbol{\mathfrak b}}^\dagger)^m\ket{\mathrm{vac}}$ and $(\boldsymbol\delta\cdot\hat{\boldsymbol{\mathfrak b}}^\dagger)^n\ket{\mathrm{vac}}$, with all quanta in one superposition of the modes, is the right-hand side of Eq.~\eqref{eq:wickgram}. We can therefore identify the function $e_{\boldsymbol\gamma}$ with the coherent state of amplitude $\boldsymbol\gamma$ and the Wick power $\wick{(\boldsymbol\gamma\cdot\mathbf z)^n}$ with the state of $n$ quanta in the mode $\boldsymbol\gamma$. The space $\mathcal H_n$ spanned by the Wick powers of degree $n$ then corresponds to the sector of $n$ quanta. The quanta are fictitious: they count the degree of a polynomial on phase space and not the photons of the field. Three consequences of this identification are used below. (i) Wick powers of different degree are orthogonal, as states with different numbers of quanta, so the spaces $\mathcal H_m$ and $\mathcal H_n$ with $m\ne n$ are orthogonal. (ii) The states with all quanta in one mode span the whole sector. This follows from the polarization identity, by which every symmetric product of $n$ vectors is a combination of the powers $\boldsymbol\gamma^{\otimes n}$. The dimension of $\mathcal H_n$ is therefore $\binom{n+2M-1}{n}$, the number of monomials of degree $n$ in $2M$ variables. It follows that $\mathcal H_n$ is the Hermite order $n$ of Sec.~\ref{sec:fisher-modes}. (iii) Since every polynomial is a sum of Wick powers and the polynomials approximate every function of finite norm, every such function is the sum of its components in the orders $n=0,1,2,\dots$.

In the notation of first quantization, the state of $n$ quanta in the mode $\boldsymbol\gamma$ is the product vector $\boldsymbol\gamma^{\otimes n}=\boldsymbol\gamma\otimes\dots\otimes\boldsymbol\gamma$. The sector of $n$ quanta consists of the vectors of $n$ factors that are symmetric under permutations of the factors. By Eq.~\eqref{eq:wickgram} the map $\boldsymbol\gamma^{\otimes n}\mapsto\wick{(\boldsymbol\gamma\cdot\mathbf z)^n}/\sqrt{n!}$ preserves the inner products. This is so because $(\bar{\boldsymbol\gamma}\cdot\boldsymbol\delta)^n=\langle\boldsymbol\gamma^{\otimes n},\boldsymbol\delta^{\otimes n}\rangle$. For a $2M\times2M$ matrix $B$, the symmetric power $\Sym^n(B)$ of Sec.~\ref{sec:fisher-modes} is the matrix $B^{\otimes n}=B\otimes\dots\otimes B$, which applies $B$ to every factor, restricted to the symmetric vectors. Hence $\Sym^n(BC)=\Sym^n(B)\Sym^n(C)$ and $\Sym^n(B)^\dagger=\Sym^n(B^\dagger)$, and for a matrix $B\ge0$ (a Hermitian matrix without negative eigenvalues) also $\Sym^n(B)\ge0$ and $\Sym^n(B)^{1/2}=\Sym^n(B^{1/2})$. The linear function $\mathbf c\cdot\mathbf x$ is the Wick power of degree one with $\boldsymbol\gamma=V^{1/2}\mathbf c$. All statements keep their form if $V^{1/2}$ is replaced by any real matrix $\Gamma$ with $\Gamma\Gamma^{\mathsf T}=V$, that is, for $\mathbf z=\Gamma^{-1}\mathbf x$, $\Lambda=\Gamma^{\mathsf T}H^{\mathsf T}(HVH^{\mathsf T}+\Sigma)^{-1}H\Gamma$ and $G_\pm=\Gamma^{-1}(V\pm i\Omega/2)(\Gamma^{\mathsf T})^{-1}$. The reason is that $\Gamma=V^{1/2}O'$ with an orthogonal matrix $O'$. This matrix only rotates the basis of the first order, and $\Sym^n(O')$ is unitary.

\emph{Rule of the coherent states.} A linear optical element, such as a beam splitter or an attenuator, maps a coherent state of amplitude $\boldsymbol\delta$ to a coherent state of amplitude $B\boldsymbol\delta$. The amplitude that all of $n$ photons pass is obtained by applying $B$ to each of them (for a lossless element this is its whole action on $n$ photons). We use the same fact for the fictitious quanta, in the following form. Let $\mathfrak B(g,h)$ be an expression that is linear in the polynomial $h$ and antilinear in the polynomial $g$, such as an inner product. Let its values on the functions $e_{s\boldsymbol\gamma}$ and $e_{t\boldsymbol\delta}$ be given term by term by the expansions~\eqref{eq:genf} in the real numbers $s$ and $t$. Moreover, let
\begin{equation}
\mathfrak B(e_{s\boldsymbol\gamma},e_{t\boldsymbol\delta})=\sum_ip_i\exp\big(st\,\bar{\boldsymbol\gamma}^{\mathsf T}B_i\boldsymbol\delta\big)
\label{eq:genform}
\end{equation}
with $2M\times2M$ matrices $B_i$ and numbers $p_i$, either for all real $\boldsymbol\gamma$ and $\boldsymbol\delta$ or for all imaginary ones (one of the two cases suffices). Then $\mathfrak B$ does not couple different orders, and in the order $n$ it is given by the matrix $\sum_ip_i\Sym^n(B_i)$, i.e., $\mathfrak B(g_n,h_n)=\sum_ip_i\langle g_n,\Sym^n(B_i)h_n\rangle$. Indeed, the coefficient of $s^mt^n$ in Eq.~\eqref{eq:genform} is $\mathfrak B\big(\wick{(\boldsymbol\gamma\cdot\mathbf z)^m},\wick{(\boldsymbol\delta\cdot\mathbf z)^n}\big)=\delta_{mn}\,n!\sum_ip_i(\bar{\boldsymbol\gamma}^{\mathsf T}B_i\boldsymbol\delta)^n$, and $(\bar{\boldsymbol\gamma}^{\mathsf T}B\boldsymbol\delta)^n=\langle\boldsymbol\gamma^{\otimes n},B^{\otimes n}\boldsymbol\delta^{\otimes n}\rangle$. Compared with Eq.~\eqref{eq:wickgram}, this is the claim for the states with all quanta in one real or one imaginary mode, and every symmetric vector is a superposition of such states.

\emph{Normal-mode basis.} We now diagonalize $J_n$. Let $O=V^{-1/2}SV_0^{1/2}$. This matrix is orthogonal, $OO^{\mathsf T}=\openone$, and it rotates the normalized quadratures of the normal modes into those of the original modes. Since $S^{-1}\Omega(S^{\mathsf T})^{-1}=\Omega$ and $V_0$ commutes with $\Omega$, the matrices of Eq.~\eqref{eq:LamG} are
\begin{equation}
G_\pm=O\big(\openone\pm\tfrac i2V_0^{-1}\Omega\big)O^{\mathsf T}.
\label{eq:Gwhite}
\end{equation}
Let $\mathbf e_\pm^{(j)}$ be $O$ applied to the vector with the entries $(1,\mp i)/\sqrt2$ in the block of the normal mode $j$ and zeros elsewhere. The functions $\mathbf e_\pm^{(j)}\cdot\mathbf z$ are, up to a factor, the phase-space functions of the creation and the annihilation operator of the normal mode $j$. These $2M$ vectors are orthonormal, with $\mathbf e_-^{(j)}=\overline{\mathbf e_+^{(j)}}$. They are also the eigenvectors of $G_\pm$: $G_\pm\mathbf e_+^{(j)}=(1\pm\tau_j)\mathbf e_+^{(j)}$ and $G_\pm\mathbf e_-^{(j)}=(1\mp\tau_j)\mathbf e_-^{(j)}$. For $\mathbf m_+=(m_{+1},\dots,m_{+M})$ and $\mathbf m_-$ with $|\mathbf m_+|+|\mathbf m_-|=n$ let $\ket{\mathbf m_+,\mathbf m_-}$ be the state with $m_{+j}$ quanta in the mode $\mathbf e_+^{(j)}$ and $m_{-j}$ quanta in the mode $\mathbf e_-^{(j)}$ for every $j$, i.e., the normalized symmetrized product of these factors. These states form an orthonormal basis of $\mathcal H_n$ (they are number states of the fictitious quanta and not Fock states of the field). Since $\Sym^n(G_\pm)$ multiplies every factor by its eigenvalue, $J_n\ket{\mathbf m_+,\mathbf m_-}=\mathsf j\ket{\mathbf m_+,\mathbf m_-}$ with
\begin{equation}
\begin{split}
\mathsf j=&\tfrac12\prod_j(1+\tau_j)^{m_{+j}}(1-\tau_j)^{m_{-j}}\\
&+\tfrac12\prod_j(1-\tau_j)^{m_{+j}}(1+\tau_j)^{m_{-j}}.
\end{split}
\label{eq:Jdiag}
\end{equation}
Since $G_-=\overline{G_+}$, the matrix $J_n$ is real in a real basis of $\mathcal H_n$. Moreover, $J_1=\openone$. All its eigenvalues are positive if no normal mode is pure. A pure mode has $\tau_j=1$, so a factor $1-\tau_j$ vanishes. Hence $\mathsf j=0$ exactly for the basis states that contain a quantum in a mode $\mathbf e_+^{(j)}$ of a pure normal mode and a quantum in a mode $\mathbf e_-^{(j')}$ of a pure normal mode, where $j'$ may differ from $j$. For a pure state and $n\ge1$ the basis states with $\mathsf j\ne0$ are the $2\binom{n+M-1}{n}$ states $\ket{\mathbf m,0}$ and $\ket{0,\mathbf m}$ with $|\mathbf m|=n$, and $\mathsf j=2^{n-1}$ for all of them.

\emph{Angular-momentum basis.} Since $O$ is a rotation for one mode, we can take $\mathbf e_\pm=(1,\mp i)^{\mathsf T}/\sqrt2$, for which $i\Omega\mathbf e_\pm=\pm\mathbf e_\pm$. Equation~\eqref{eq:Gwhite} then reads $G_\pm=\openone\pm i\tau\,\Omega$ for every $V$. We write $\ket k$, $k=0,\dots,n$, for the state with $n-k$ quanta in the mode $\mathbf e_+$ and $k$ quanta in the mode $\mathbf e_-$, and call these states the angular-momentum basis of $\mathcal H_n$. Equation~\eqref{eq:Jdiag} is then Eq.~\eqref{eq:Jeig}, and for $v=1/2$ and $n\ge2$ only $\ket0$ and $\ket n$ have a nonzero eigenvalue. Under a rotation $R_\varphi$ of phase space by the angle $\varphi$ the vectors $\mathbf e_\pm$ are multiplied by $e^{\pm i\varphi}$, and hence the state $\ket k$ by $e^{i(n-2k)\varphi}$. Hence $\ket k$ is the mode of angular momentum $l=n-2k$, and the coefficients of $g_n$ in this basis are the $b_{n,l}$ of Sec.~\ref{sec:fisher-thermal}. The opposite sign convention for $\Omega$ exchanges $k$ and $n-k$, and all statements of the paper are invariant under this exchange.

\subsection{Proof of Theorem~\ref{thm:structure}}\label{app:structure}
The proof has three steps. In the first two steps we evaluate the classical and the quantum Fisher information on the coherent states of Appendix~\ref{app:conv} and apply to them the rule of Eq.~\eqref{eq:genform}. In this way we obtain the theorem for all directions that are described by polynomials. In the third step we extend it to all directions of finite quantum Fisher information.

\emph{Step 1: measurement.} Let $g=W_\Delta/W_0$ have a finite norm. We describe the measurement by two random vectors. One of them is a phase-space point $\mathbf X$ distributed with $W_0$, and the other is the outcome $\mathbf Y=H\mathbf X+\mathbf N$. The noise $\mathbf N$ is normal, with zero mean and the covariance matrix $\Sigma$, and it is independent of $\mathbf X$. We claim that the ratio of the outcome densities is the average of $g$ over the points that give the outcome, $\delta(\mathbf y)/f(\mathbf y)=\mathbb E[g(\mathbf X)|\mathbf Y=\mathbf y]$ (the conditional expectation). Here $\mathbb E$ is the average over $\mathbf X$ and $\mathbf N$, whereas in the proof of Theorem~\ref{thm:mix} it is the average over the settings. We also claim that
\begin{equation}
F(\Delta)=\sum_n\langle g_n,\Sym^n(\Lambda)g_n\rangle\le\|g\|^2 .
\label{eq:classblocks}
\end{equation}
To show the first claim we compare characteristic functions. Since the outcome is a combination of quadratures plus independent noise, its characteristic function for an operator $A$ in place of the state is the product of those of the two terms. As a function of $\mathbf q\in\mathbb R^d$ it reads $\tr(Ae^{i\mathbf q\cdot H\hat{\mathbf x}})e^{-\mathbf q^{\mathsf T}\Sigma\mathbf q/2}$. For $A=\Delta$ we have $\tr(\Delta D_{\mathbf b})=\int gW_0e^{i\mathbf b\cdot\mathbf x}$, and the characteristic function is then $\mathbb E[g(\mathbf X)e^{i\mathbf q\cdot \mathbf Y}]$. Averaging first over $\mathbf X$ at a fixed outcome, we write it as $\mathbb E\{\mathbb E[g(\mathbf X)|\mathbf Y]e^{i\mathbf q\cdot \mathbf Y}\}$. This is the characteristic function of the density $\delta=f\,\mathbb E[g(\mathbf X)|\mathbf Y]$, where $f$ is the outcome density of $\rho_0$. Hence $F(\Delta)=\int\delta^2/f=\mathbb E\{\mathbb E[g|\mathbf Y]^2\}$. Since the square of an average is at most the average of the square, $F(\Delta)\le\mathbb E[g(\mathbf X)^2]=\|g\|^2$. By the same fact and the Cauchy--Schwarz inequality, the expression $\mathfrak B(g,h)=\mathbb E\{\overline{\mathbb E[g|\mathbf Y]}\,\mathbb E[h|\mathbf Y]\}$ obeys $|\mathfrak B(g,h)|\le\|g\|\,\|h\|$. By Eq.~\eqref{eq:wickgram} the norm of the $n$th term of the series of Eq.~\eqref{eq:genf} is $|t|^n|\boldsymbol\gamma|^n/\sqrt{n!}$, so the series converges in the norm. This allows us to evaluate $\mathfrak B$ on it term by term, and a relation that holds for all polynomials holds for all functions of finite norm.

We now evaluate $\mathfrak B$ on the coherent states. Let $\Xi=(HVH^{\mathsf T}+\Sigma)^{-1}$, the inverse of the covariance matrix of the outcome. If $HVH^{\mathsf T}+\Sigma$ has no inverse, a combination of the outcomes vanishes for every state and we drop it. The distribution of $\mathbf X$ at a given outcome $\mathbf y$ follows from the standard formulas for normal variables (they are also the basis of linear filtering). It is normal, with the mean $VH^{\mathsf T}\Xi\mathbf y$ and the covariance matrix $V-VH^{\mathsf T}\Xi HV$. The average of the exponential function $e_{\boldsymbol\gamma}$ over this distribution is again an exponential function. With $\tilde{\boldsymbol\gamma}=V^{-1/2}\boldsymbol\gamma$ and $\tilde{\boldsymbol\delta}=V^{-1/2}\boldsymbol\delta$,
\begin{equation}
\mathbb E[e_{\boldsymbol\gamma}(\mathbf X)|\mathbf Y=\mathbf y]=\exp\big(\tilde{\boldsymbol\gamma}^{\mathsf T}VH^{\mathsf T}\Xi\,\mathbf y-\tfrac12\tilde{\boldsymbol\gamma}^{\mathsf T}VH^{\mathsf T}\Xi HV\tilde{\boldsymbol\gamma}\big),
\label{eq:condexp}
\end{equation}
and by averaging the product of two such functions over the outcome we obtain $\mathfrak B(e_{\boldsymbol\gamma},e_{\boldsymbol\delta})=\exp(\bar{\tilde{\boldsymbol\gamma}}^{\mathsf T}VH^{\mathsf T}\Xi HV\tilde{\boldsymbol\delta})=\exp(\bar{\boldsymbol\gamma}^{\mathsf T}\Lambda\boldsymbol\delta)$ with the matrix $\Lambda=V^{1/2}H^{\mathsf T}\Xi HV^{1/2}$ of Eq.~\eqref{eq:LamG}. In the language of the analogy, the measurement maps a coherent state of amplitude $\boldsymbol\gamma$ to a coherent state, as an attenuator does. The matrix $\Lambda$ then plays the role of its intensity transmission, and with the rule of Eq.~\eqref{eq:genform} we obtain the matrices $\Sym^n(\Lambda)$ of Eq.~\eqref{eq:classblocks}.

In this derivation we used only $\Sigma\ge0$. Condition~\eqref{eq:phys} guarantees in addition that the distribution of $\mathbf Y$ is that of a measurement, and this property is needed in Step 3. To see this, we add an ancilla with the operators $\hat{\mathbf N}$ and consider the $d$ operators $H\hat{\mathbf x}\otimes\openone+\openone\otimes\hat{\mathbf N}$. They commute with each other if $[\hat N_i,\hat N_{i'}]=-i(H\Omega H^{\mathsf T})_{ii'}$. This is because the commutators of the two terms then cancel (the components of $\hat{\mathbf N}$ that commute with all others are classical random variables). Condition~\eqref{eq:phys} is the uncertainty relation for operators with these commutators. If it holds, the ancilla has a Gaussian state in which $\hat{\mathbf N}$ has zero mean and the covariance matrix $\Sigma$, and the joint measurement of the $d$ commuting operators in this state has the distribution of $\mathbf Y$. For heterodyne detection of one mode this is the known scheme in which the mode is mixed with a vacuum ancilla on a balanced beam splitter and $\hat x$ is measured at one output and $\hat p$ at the other.

\emph{Step 2: quantum Fisher information.} We first write the quantum Fisher information as the squared norm of the SLD. For operators $C$ and $Z$ with the elements $C_{ij}$ and $Z_{ij}$ in the eigenbasis of $\rho_0$ let
\begin{equation}
\begin{split}
\langle C,Z\rangle_{\mathfrak L}&=\tfrac12\tr\big[\rho_0\big(C^\dagger Z+ZC^\dagger\big)\big]\\
&=\sum_{ij}\tfrac12(\pi_i+\pi_j)\,\overline{C_{ij}}Z_{ij},
\end{split}
\label{eq:Linner}
\end{equation}
the symmetrized correlation function of $C^\dagger$ and $Z$ in the state $\rho_0$, with the norm $\|Z\|^2_{\mathfrak L}=\langle Z,Z\rangle_{\mathfrak L}$. We denote by $\mathfrak L$ the space of the operators of finite norm (a Hilbert space). In this space two operators count as equal if they agree in all elements with $\pi_i+\pi_j>0$. This is possible since the other elements do not enter the norm. Let $\mathcal J(Z)=(\rho_0Z+Z\rho_0)/2$, the operator with the elements $\tfrac12(\pi_i+\pi_j)Z_{ij}$. The SLD $L$ of a direction is defined by $\Delta=\mathcal J(L)$, and Eq.~\eqref{eq:fsld} states that $F_{\mathrm{SLD}}(\Delta)$ is finite if and only if a Hermitian $L\in\mathfrak L$ with $\Delta=\mathcal J(L)$ exists, and that then $F_{\mathrm{SLD}}(\Delta)=\|L\|^2_{\mathfrak L}$.

Let $A$ and $C$ be polynomial operators with the phase-space functions $w_A$ and $w_C$, and let $\mathsf J$ be the operation that multiplies the component of order $n$ of a polynomial by $J_n$. We claim that
\begin{equation}
\begin{split}
\tr[C^\dagger\mathcal J(A)]&=\langle C,A\rangle_{\mathfrak L}=\langle w_C,\mathsf Jw_A\rangle,\\
W_{\mathcal J(A)}&=W_0\,\mathsf Jw_A .
\end{split}
\label{eq:quantblocks}
\end{equation}
The first relation says that the correlation function of two polynomial operators is given, order by order, by the matrices $J_n$. According to the second one, the direction with the SLD $A$ has the modes $g_n=J_n(w_A)_n$. For the proof we use the displacement operators as the coherent states of the analogy. The phase-space function of $D_{\mathbf a}$ is $e^{i\mathbf a\cdot\mathbf x}=e^{-\mathbf a^{\mathsf T}V\mathbf a/2}e_{\boldsymbol\gamma}$ with the imaginary amplitude $\boldsymbol\gamma=iV^{1/2}\mathbf a$. Hence $e_{t\boldsymbol\gamma}$ is the phase-space function of $e^{t^2\mathbf a^{\mathsf T}V\mathbf a/2}D_{t\mathbf a}$. The terms of its expansion in $t$, i.e., the Wick powers, are then the phase-space functions of the terms in the expansion of this operator, and these terms are polynomial operators. By the composition law and the characteristic function of $\rho_0$, for real $\mathbf a$ and $\mathbf c$,
\begin{equation}
\tr\big(D_{\mathbf c}^\dagger\rho_0D_{\mathbf a}\big)=e^{i\mathbf a^{\mathsf T}\Omega\mathbf c/2}\,e^{-(\mathbf a-\mathbf c)^{\mathsf T}V(\mathbf a-\mathbf c)/2},
\label{eq:gen}
\end{equation}
and for the other order of the operators, $\tr(D_{\mathbf c}^\dagger D_{\mathbf a}\rho_0)$, we obtain the complex conjugate. The phase factor comes from the commutator of the quadratures and has opposite signs for the two orders. With $\boldsymbol\gamma'=iV^{1/2}\mathbf c$ the average of the two traces, multiplied by $e^{(\mathbf a^{\mathsf T}V\mathbf a+\mathbf c^{\mathsf T}V\mathbf c)/2}$, is
\begin{equation}
\tfrac12\sum_\pm\exp\big[\mathbf c^{\mathsf T}\big(V\pm\tfrac i2\Omega\big)\mathbf a\big]=\tfrac12\sum_\pm\exp\big(\bar{\boldsymbol\gamma}'^{\mathsf T}G_\pm\boldsymbol\gamma\big).
\label{eq:bopp}
\end{equation}
This is the form of Eq.~\eqref{eq:genform} with the two matrices $G_+$ and $G_-$ and the weights $1/2$. The symmetrized correlation function thus treats the coherent states as two linear elements would, one for each order of the operators, and the matrices $G_\pm$ of the uncertainty relation play the role of their transmissions. However, the analogy is formal here: the eigenvalues $1\pm\tau_j$ of $G_\pm$ are not bounded by one, so each element attenuates one mode of every pair $\mathbf e_\pm^{(j)}$ and amplifies the other. Applying the rule of Eq.~\eqref{eq:genform} to the expression $(w_C,w_A)\mapsto\tr[C^\dagger\mathcal J(A)]$, we obtain the first relation of Eq.~\eqref{eq:quantblocks} with $J_n=[\Sym^n(G_+)+\Sym^n(G_-)]/2$.

\emph{Technical point.} The rule requires that the traces can be expanded term by term in $s$ and $t$. We write $\tr(D_{s\mathbf c}^\dagger\rho_0D_{t\mathbf a})=\sum_{\mathbf m}\pi_{\mathbf m}\langle D_{t\mathbf a}^\dagger\Psi_{\mathbf m},D_{s\mathbf c}^\dagger\Psi_{\mathbf m}\rangle$ and expand the displaced eigenvectors in $t$ and $s$. Since $\mathbf a\cdot\hat{\mathbf x}$ is a sum of an annihilation and a creation operator of the normal modes, $\|(\mathbf a\cdot\hat{\mathbf x})^n\ket{\Psi_{\mathbf m}}\|\le(c_S|\mathbf a|)^n[(|\mathbf m|+n)!/|\mathbf m|!]^{1/2}$ with a constant $c_S$ that depends on $S$ only. Hence the expansion of $D_{t\mathbf a}\ket{\Psi_{\mathbf m}}$ in $t$ converges, and for every $\varrho'<1$ the sum of the norms of its terms is at most a constant times $\varrho'^{-|\mathbf m|/2}$. The thermal weights fall faster, $\pi_{\mathbf m}\le\varrho^{|\mathbf m|}$ with $\varrho=\max_jq_j<1$, and $\varrho'$ can be chosen larger than $\varrho$. The series in $s$, $t$ and $\mathbf m$ therefore converges absolutely, which permits the expansion under the trace. The same holds for the other order of the operators, and the average of the two traces of the terms is $\langle C,A\rangle_{\mathfrak L}$.

\emph{Step 2, continued.} To obtain the second relation of Eq.~\eqref{eq:quantblocks}, we take the Fourier transform of $\tr(\rho_0D_{\mathbf a}D_{\mathbf b})$ in $\mathbf b$. It shows that $W_{\rho_0D_{\mathbf a}}/W_0$ is the exponential of a linear function of $\mathbf x$, so $W_{\rho_0A}/W_0$ and $W_{A\rho_0}/W_0$ are polynomials of the degree of $w_A$. By differentiating $\tr[\mathcal J(A)D_{t\mathbf b}]=\int W_{\mathcal J(A)}e^{it\mathbf b\cdot\mathbf x}$ in $t$ we find $\tr[C^\dagger\mathcal J(A)]=\langle w_C,W_{\mathcal J(A)}/W_0\rangle$ for every polynomial operator $C$. Comparing this with the first relation, we see that the polynomial $W_{\mathcal J(A)}/W_0$ equals $\mathsf Jw_A$.

In summary, for a Hermitian polynomial operator $L$ with $\tr(\rho_0L)=0$ we obtain from Eq.~\eqref{eq:quantblocks} a direction $\Delta=\mathcal J(L)$ with the polynomial $g=\mathsf Jw_L$ and $F_{\mathrm{SLD}}(\Delta)=\langle w_L,\mathsf Jw_L\rangle=\sum_n\langle g_n,J_n^{-1}g_n\rangle$. Its classical Fisher information follows from Step 1. In this way we have proved Theorem~\ref{thm:structure} for the directions whose SLD is a polynomial in the quadratures. Note that $\mathsf J$ is applied here to polynomials only, and it cannot be applied to all functions of finite norm, because the largest eigenvalue of $J_n$ grows with $n$ at least as $(1+\max_j\tau_j)^n/2$.

\emph{Step 3: extension to all directions.} A general SLD is not a polynomial in the quadratures, e.g., that of a cat state admixed to a thermal state. We show that it is a limit of polynomial operators and that both Fisher informations are continuous in this limit. For this we need two properties of $\mathfrak L$. (i) For every measurement and every Hermitian $Z\in\mathfrak L$, the direction $\mathcal J(Z)$ has an outcome density $\delta$ with $\int\delta^2/f\le\|Z\|^2_{\mathfrak L}$. This is the inequality $F\le F_{\mathrm{SLD}}$ of Braunstein and Caves~\citep{BraunsteinCaves1994}, stated for all operators of $\mathfrak L$. (ii) Every operator of $\mathfrak L$ can be approximated in the norm of $\mathfrak L$ by polynomial operators. Polynomial operators belong to $\mathfrak L$ because the thermal weights $\pi_{\mathbf m}$ fall exponentially with the photon number, whereas the elements of a polynomial operator grow only as a power of it. The same exponential decay excludes an operator of $\mathfrak L$ that is orthogonal to all of them.

\emph{Technical point: proofs of (i) and (ii).} (i) With $R_{ij}=\pi_i^{1/2}Z_{ij}$ and $R'_{ij}=Z_{ij}\pi_j^{1/2}$ we have $\mathcal J(Z)=(\rho_0^{1/2}R+R'\rho_0^{1/2})/2$ and $\tr(R^\dagger R)+\tr(R'^\dagger R')=2\|Z\|^2_{\mathfrak L}$. For a bounded real function $\varkappa$ of the outcome let $\hat\varkappa$ be the operator of the measurement whose mean is the average of $\varkappa$, $\tr(\rho\hat\varkappa)=\int\varkappa f_\rho$ for every state $\rho$. The decomposition of $\mathcal J(Z)$, the Cauchy--Schwarz inequality and $\tr(\rho_0\hat\varkappa^2)\le\tr(\rho_0\widehat{\varkappa^2})$ give $|\int\varkappa\,\delta|\le\|Z\|_{\mathfrak L}(\int\varkappa^2f)^{1/2}$. A linear expression in $\varkappa$ with this bound is the integral of $\varkappa$ with a density $\delta$ that obeys $\int\delta^2/f\le\|Z\|^2_{\mathfrak L}$ (the Riesz theorem), which is the claim. (ii) A Gaussian unitary operation maps polynomial operators to polynomial operators and preserves the inner product of $\mathfrak L$. By Appendix~\ref{app:conv} we can therefore take $\rho_0$ to be the product of the thermal states of the normal modes, with the Fock basis $\ket{\mathbf m}$. Let $Z\in\mathfrak L$ be orthogonal to all polynomial operators. We show that $Z=0$, which is equivalent to (ii). Let $\boldsymbol\ell$ be a vector of $M$ integers. For a polynomial $P$ in $M$ variables the operator $A=P(\nhat_1,\dots,\nhat_M)\prod_jT_j$, with $T_j=\hat a_j^{\ell_j}$ for $\ell_j\ge0$ and $T_j=(\hat a_j^\dagger)^{|\ell_j|}$ for $\ell_j<0$, is polynomial, and its only nonzero elements are $A_{\mathbf m,\mathbf m+\boldsymbol\ell}=P(\mathbf m)\,c_{\mathbf m}$ with positive numbers $c_{\mathbf m}$ that grow at most as a power of $|\mathbf m|$. The orthogonality reads $\sum_{\mathbf m}\overline{P(\mathbf m)}\,u_{\mathbf m}=0$ with $u_{\mathbf m}=\varpi_{\mathbf m}c_{\mathbf m}Z_{\mathbf m,\mathbf m+\boldsymbol\ell}$ and the weight $\varpi_{\mathbf m}=(\pi_{\mathbf m}+\pi_{\mathbf m+\boldsymbol\ell})/2$. This weight vanishes unless the photon numbers of the pure modes vanish in $\mathbf m$ or in $\mathbf m+\boldsymbol\ell$, so $\varpi_{\mathbf m}\le c\,\varrho_1^{|\mathbf m|}$ with a constant $c$ and $\varrho_1=\max(1/2,\max_jq_j)$. By the Cauchy--Schwarz inequality $\sum_{\mathbf m}|u_{\mathbf m}|e^{\mathbf s\cdot\mathbf m}\le(\sum_{\mathbf m}\varpi_{\mathbf m}c_{\mathbf m}^2e^{2\mathbf s\cdot\mathbf m})^{1/2}\|Z\|_{\mathfrak L}$, which is finite for real $\mathbf s$ with $\max_js_j<\tfrac12\ln(1/\varrho_1)$. The function $\sum_{\mathbf m}u_{\mathbf m}e^{\mathbf s\cdot\mathbf m}$ of $\mathbf s\in\mathbb C^M$ is therefore analytic in a connected region that contains $\mathbf s=0$ and the imaginary axes. By the orthogonality to all polynomials $P$ all its derivatives vanish at $\mathbf s=0$, so it vanishes everywhere. For imaginary $\mathbf s$ it is a Fourier series with $\sum_{\mathbf m}|u_{\mathbf m}|<\infty$. All its coefficients then vanish, $u_{\mathbf m}=0$, and $Z_{\mathbf m,\mathbf m+\boldsymbol\ell}=0$ whenever $\varpi_{\mathbf m}>0$. Since $\boldsymbol\ell$ is arbitrary, $Z=0$ in $\mathfrak L$.

\emph{Precise form of Theorem~\ref{thm:structure}.} Let $F_{\mathrm{SLD}}(\Delta)<\infty$. (a) All polynomial moments $\tr[\mathrm{Op}(h)^\dagger\Delta]$ of $\Delta$ exist, there are unique real $g_n\in\mathcal H_n$ with $\langle h,g_n\rangle=\tr[\mathrm{Op}(h)^\dagger\Delta]$ for all $h\in\mathcal H_n$, every $g_n$ lies in the range of $J_n$, and Eqs.~\eqref{eq:structure} and~\eqref{eq:blocks} hold, with $J_n^{-1}$ the inverse on the range. (b) If $g=W_\Delta/W_0$ has a finite norm, the $g_n$ are its components in the Hermite orders.

\emph{Proof.} (a) Let $\Delta=\mathcal J(L)$ and, by property (ii), let $L^{(\iota)}$, $\iota=1,2,\dots$, be Hermitian polynomial operators that converge to $L$ in $\mathfrak L$ (if polynomial operators converge to the Hermitian $L$, so do their Hermitian parts), with the directions $\Delta^{(\iota)}=\mathcal J(L^{(\iota)})$. By Eq.~\eqref{eq:quantblocks} the modes of $\Delta^{(\iota)}$ are $g_n^{(\iota)}=J_n^{1/2}h_n^{(\iota)}$, where $h_n^{(\iota)}$ is $J_n^{1/2}$ applied to the component of order $n$ of the phase-space function of $L^{(\iota)}$, and $\|L^{(\iota)}-L^{(\iota')}\|^2_{\mathfrak L}=\sum_n\|h_n^{(\iota)}-h_n^{(\iota')}\|^2$. Hence $h_n^{(\iota)}$ converges for every $n$ to a vector $h_n$ in the range of $J_n$, with $\sum_n\|h_n\|^2=\|L\|^2_{\mathfrak L}=F_{\mathrm{SLD}}(\Delta)$. The vector $g_n=J_n^{1/2}h_n$ then obeys the second relation of Eq.~\eqref{eq:structure}. For a polynomial operator $A$, the sum for $\tr(A^\dagger\Delta)$ converges absolutely by the Cauchy--Schwarz inequality in $\mathfrak L$, and
\begin{equation}
\begin{split}
\tr(A^\dagger\Delta)&=\langle A,L\rangle_{\mathfrak L}=\lim_\iota\langle A,L^{(\iota)}\rangle_{\mathfrak L}\\
&=\sum_n\langle(w_A)_n,g_n\rangle,
\end{split}
\label{eq:momentlimit}
\end{equation}
which is the statement about the moments. For the classical Fisher information we obtain from Eq.~\eqref{eq:classblocks} $F(\Delta^{(\iota)})=\sum_n\langle h_n^{(\iota)},E_nh_n^{(\iota)}\rangle$ with $E_n=J_n^{1/2}M_nJ_n^{1/2}$. By property (i), no eigenvalue of $E_n$ exceeds one. By the same property the ratios $\delta/f$ of $\Delta^{(\iota)}$ converge to that of $\Delta$ in the mean square, i.e., $\int(\delta^{(\iota)}-\delta)^2/f\to0$ for their outcome densities $\delta^{(\iota)}$, so
\begin{equation}
\begin{split}
F(\Delta)&=\lim_\iota F(\Delta^{(\iota)})=\sum_n\langle h_n,E_nh_n\rangle\\
&=\sum_n\langle g_n,M_ng_n\rangle .
\end{split}
\label{eq:Flimit}
\end{equation}
For a random setting that is recorded, $F$ and therefore also $M_n$ are averaged over the settings. (b) Let $\tilde g_n$ be the components of $g$ in the Hermite orders. For heterodyne detection of every mode the ratio $\delta/f$ of $\Delta$ is the image of $g$ under the conditional expectation $g\mapsto\mathbb E[g(\mathbf X)|\mathbf Y]$ of Step 1 and the mean-square limit of the images of the polynomials $\sum_ng_n^{(\iota)}$, so $\sum_n\langle\tilde g_n-g_n^{(\iota)},\Sym^n(\Lambda)(\tilde g_n-g_n^{(\iota)})\rangle\to0$. Since all eigenvalues of $\Lambda$ are positive for heterodyne detection, $g_n^{(\iota)}\to\tilde g_n$ for every $n$, and therefore $\tilde g_n=g_n$.

\emph{Efficiencies.} We write $g_n=J_n^{1/2}h_n$ with $h_n$ in the range of $J_n$. Then $\langle g_n,J_n^{-1}g_n\rangle=\|h_n\|^2$ and $\langle g_n,M_ng_n\rangle=\langle h_n,E_nh_n\rangle$ with $E_n=J_n^{1/2}M_nJ_n^{1/2}$. The efficiency in the order $n$ is thus the mean value of $E_n$ in the normalized vector $h_n/\|h_n\|$, as the mean energy of a state is the mean value of the Hamiltonian. Its stationary values are the eigenvalues of $E_n$ on the range of $J_n$, and its largest value is
\begin{equation}
\sup\frac{\langle g_n,M_ng_n\rangle}{\langle g_n,J_n^{-1}g_n\rangle}=\lambda_{\max}(E_n)=\lambda_{\max}\big(M_n^{1/2}J_nM_n^{1/2}\big).
\label{eq:effsup}
\end{equation}
The second equality holds because the product of two matrices has the same nonzero eigenvalues in either order, here for $J_n^{1/2}M_n^{1/2}$ and $M_n^{1/2}J_n^{1/2}$. For several orders $F/F_{\mathrm{SLD}}=\sum_n\langle h_n,E_nh_n\rangle/\sum_n\|h_n\|^2$. This ratio is the average of the efficiencies $\langle h_n,E_nh_n\rangle/\|h_n\|^2$ of the orders, weighted with their quantum Fisher informations $\|h_n\|^2$. $\square$

For a pure reference state $\ket{\Psi_0}$, the eigenvector with $\mathbf m=0$, only $\pi_0=1$ is nonzero. Equation~\eqref{eq:fsld} is then finite if and only if $\Delta$ has nonzero elements only in the row and the column of $\ket{\Psi_0}$, i.e., $\Delta=\ket{\Psi_0}\!\bra\psi+\ket\psi\!\bra{\Psi_0}$ with $\braket{\Psi_0|\psi}=0$, and then $F_{\mathrm{SLD}}=4\braket{\psi|\psi}=\sum_{n\ge1}2^{1-n}\|g_n\|^2$.

\subsection{Proofs of the bounds}\label{app:bounds}
The proofs rest on two inequalities between matrices. For Hermitian matrices $A\le B$ means that $B-A$ has no negative eigenvalue, i.e., that the mean value of $A$ does not exceed that of $B$ in any vector.

\emph{Attenuation of all quanta.} We start with the first inequality, which concerns $n$ quanta. For a matrix $0\le B\le\openone$ and $n\ge1$,
\begin{equation}
B^{\otimes n}\le\frac1n\sum_{i=1}^nB^{(i)},
\label{eq:tensor}
\end{equation}
where $B^{(i)}$ acts as $B$ on the $i$th factor and as the identity on the others. If $B$ is the transmission of an attenuator, this inequality says that the probability that all $n$ quanta pass does not exceed the probability that a given one passes. For the proof we note that $B^{(1)}-B^{\otimes n}=B\otimes(\openone-B^{\otimes(n-1)})\ge0$. Since the same holds for every position $i$, it holds for the average over $i$.

\emph{Two uncertainty relations.} In the second inequality we combine the uncertainty relation of the state with that of the measurement. For one setting that obeys Eq.~\eqref{eq:phys} let $B_\pm=\Lambda^{1/2}G_\pm\Lambda^{1/2}$ and $C_\pm=G_\pm^{1/2}\Lambda G_\pm^{1/2}$. Then
\begin{equation}
0\le B_\pm\le\openone,\qquad0\le C_\pm\le\openone,\qquad\tfrac12(B_++B_-)=\Lambda .
\label{eq:uncert}
\end{equation}
Indeed, $G_\pm\ge0$ is the uncertainty relation of the state, and the last relation follows from $(G_++G_-)/2=\openone$. For the real matrix $\Upsilon=\Xi^{1/2}HV^{1/2}$ we have $\Lambda=\Upsilon^{\mathsf T}\Upsilon$. The matrices $B_\pm$, $C_\pm$ and $\Upsilon G_\pm\Upsilon^{\mathsf T}$ then have the same nonzero eigenvalues, and
\begin{equation}
\begin{split}
\Upsilon G_\pm\Upsilon^{\mathsf T}&=\Xi^{1/2}H\big(V\pm\tfrac i2\Omega\big)H^{\mathsf T}\Xi^{1/2}\\
&=\openone-\Xi^{1/2}\big(\Sigma\mp\tfrac i2H\Omega H^{\mathsf T}\big)\Xi^{1/2}\le\openone,
\end{split}
\label{eq:XGX}
\end{equation}
where in the second line we used $HVH^{\mathsf T}=\Xi^{-1}-\Sigma$, and the inequality is the uncertainty relation~\eqref{eq:phys} of the measurement. Thus $B_\pm$ and $C_\pm$ are Hermitian matrices between $0$ and $\openone$, and Eq.~\eqref{eq:tensor} applies to them as to the transmission of an attenuator. For homodyne detection of commuting quadratures $\Lambda$ is the projector onto the measured combinations of the normalized quadratures, i.e., onto the vectors $V^{1/2}H^{\mathsf T}\mathbf q$. In this case $H\Omega H^{\mathsf T}=0$, which gives $\Lambda V^{-1/2}\Omega V^{-1/2}\Lambda=0$ and $B_\pm=\Lambda$.

\emph{Proof of Theorem~\ref{thm:single}.} By Eq.~\eqref{eq:effsup} and the product rule of $\Sym^n$, the largest efficiency in the order $n$ is the largest eigenvalue of
\begin{equation}
M_n^{1/2}J_nM_n^{1/2}=\tfrac12\big[\Sym^n(B_+)+\Sym^n(B_-)\big].
\label{eq:MJM}
\end{equation}
By Eqs.~\eqref{eq:tensor} and~\eqref{eq:uncert} the mean value of this matrix in any symmetric vector is at most that of $\frac1n\sum_i[(B_++B_-)/2]^{(i)}=\frac1n\sum_i\Lambda^{(i)}$, which is at most $\lambda_{\max}(\Lambda)$. We pass from $n$ quanta to one quantum with the inequality~\eqref{eq:tensor}, and for one quantum the two matrices $B_\pm$ average to $\Lambda$. Hence no efficiency of any order exceeds $\lambda_{\max}(\Lambda)$. Since the efficiency of a direction is an average over its orders, we obtain Eq.~\eqref{eq:thmsingle}. In the first order $J_1=\openone$ and $E_1=\Lambda$. The bound is then attained by the eigenvector $\mathbf c$ of $\Lambda$ with the largest eigenvalue, i.e., by the mode $g(\mathbf x)=\mathbf c\cdot V^{-1/2}\mathbf x$. A displacement by $\varepsilon\mathbf x_0$ changes $W_0$ by $-\varepsilon\,\mathbf x_0\cdot\nabla W_0$ and therefore has $g(\mathbf x)=\mathbf x_0\cdot V^{-1}\mathbf x$, which coincides with the mode above for $\mathbf x_0=V^{1/2}\mathbf c$. For homodyne detection of $d$ commuting quadratures $B_\pm=\Lambda$ is a projector onto a space of dimension $d$. The matrix~\eqref{eq:MJM} is then $\Sym^n(\Lambda)$, the projector onto the states of $n$ quanta in these $d$ modes, and the number of these states is $\binom{n+d-1}{n}$. Thus $E_n$ has the eigenvalue $1$ with this multiplicity and no other nonzero eigenvalue. Its eigenvectors with the eigenvalue $1$ are $J_n^{1/2}w$, where $w$ is a state of $n$ quanta in the $d$ modes. They belong to the directions with $g_n=J_nw$, and the SLDs $\mathrm{Op}(w)$ of these directions are polynomials of degree $n$ in the measured quadratures. For one mode $d=1$, $\mathbf c\propto V^{1/2}\mathbf u$ and $\mathbf x_0\propto V\mathbf u$. $\square$

Let us show that condition~\eqref{eq:phys} cannot be dropped. If it fails, by Eq.~\eqref{eq:XGX} $B_+$ or $B_-$ has an eigenvalue $b>1$, and the matrix~\eqref{eq:MJM} has an eigenvalue of at least $b^n/2$, which exceeds one for large $n$. A direction with a polynomial SLD would then have $F>F_{\mathrm{SLD}}$. Since no measurement allows this, the distribution of $\mathbf Y$ does not belong to a measurement. As an example, we take one mode with $V=\openone/2$, $H=\openone$ and $\Sigma=0.1\,\openone$, i.e., a simultaneous measurement of $\hat x$ and $\hat p$ with less added noise than the uncertainty relation requires. In this case we find $\lambda_{\max}(\Lambda)=5/6$ and $\lambda_{\max}(E_2)=25/18$.

\emph{Proof of Theorem~\ref{thm:mix}.} Here $M=1$, and we use the angular-momentum basis $\ket k$ of Appendix~\ref{app:conv}. We first bound the diagonal elements of the efficiency matrix of one setting and then find how its other elements depend on the phase of the setting. At the end we average over the settings.

First, for one setting
\begin{equation}
\bra kE_n\ket k\le\tfrac12\tr\Lambda\qquad(k=0,\dots,n) .
\label{eq:diag}
\end{equation}
The matrix $J_n$ is diagonal in the basis $\ket k$, so
\begin{equation}
\begin{split}
\bra kE_n\ket k&=\tfrac12\sum_\pm\bra k\Sym^n(C_\pm)\ket k\\
&\le\tfrac12\sum_\pm\frac1n\sum_i\bra kC_\pm^{(i)}\ket k
\end{split}
\label{eq:diagstep}
\end{equation}
by Eqs.~\eqref{eq:tensor} and~\eqref{eq:uncert}. In the last expression only the state of a single quantum appears. In $\ket k$ a given quantum is in the mode $\mathbf e_+$ with the probability $(n-k)/n$ and in the mode $\mathbf e_-$ with the probability $k/n$, without coherence between the two. Its state is then $\frac{n-k}n\ket{\mathbf e_+}\!\bra{\mathbf e_+}+\frac kn\ket{\mathbf e_-}\!\bra{\mathbf e_-}$. For the real symmetric matrix $\Lambda$, $\bra{\mathbf e_\pm}\Lambda\ket{\mathbf e_\pm}=\tr\Lambda/2$. Since $\mathbf e_\pm$ are eigenvectors of $G_\pm$, we have $\bra{\mathbf e_+}C_\pm\ket{\mathbf e_+}=(1\pm\tau)\tr\Lambda/2$ and $\bra{\mathbf e_-}C_\pm\ket{\mathbf e_-}=(1\mp\tau)\tr\Lambda/2$. In the average over the two signs $\tau$ drops out, and we obtain Eq.~\eqref{eq:diag}.

Second, let $\Lambda_0=\mathrm{diag}(\mu_+,\mu_-)$ be the matrix of the setting with the phase $0$. Since $G_\pm=\openone\pm i\tau\,\Omega$ commute with the rotations of the normalized quadratures, the matrices $B_\pm$ and $C_\pm$ of $\Lambda_0$ have the same eigenvalues as those of the original setting. This allows us to apply Eq.~\eqref{eq:uncert} and Theorem~\ref{thm:single} to $\Lambda_0$. In the basis $\mathbf e_\pm$ its diagonal elements are $(\mu_++\mu_-)/2$ and its off-diagonal elements are $(\mu_+-\mu_-)/2$, all nonnegative. Hence $\Sym^n(\Lambda_0)$ and the efficiency matrix $E^{(0)}$ of this setting have nonnegative elements in the angular-momentum basis. For the setting with the phase $\varphi$ we have $\Lambda=R_\varphi\Lambda_0R_\varphi^{\mathsf T}$. The rotation multiplies $\ket k$ by $e^{i(n-2k)\varphi}$, which gives $\bra kE^{(\varphi)}\ket{k'}=e^{2i(k'-k)\varphi}E^{(0)}_{kk'}$. The phase of the setting thus enters only through phase factors of the elements between different angular momenta.

Third, we average over the phase and, independently, over the eigenvalues,
\begin{equation}
(E_n)_{kk'}=\bar E_{kk'}\,\hat w_{k'-k},\qquad\bar E=\mathbb E\,E^{(0)}.
\label{eq:Emix}
\end{equation}
The diagonal elements are not affected by the average over the phase, and the others are reduced by the Fourier coefficients of its distribution. The matrix $\bar E$ has nonnegative elements. Its diagonal elements obey $\bar E_{kk}\le\mu_{\mathrm{iso}}$ by Eq.~\eqref{eq:diag}, and by Theorem~\ref{thm:single} $\lambda_{\max}(\bar E)\le\mathbb E\mu_+=\mu_{\mathrm{iso}}+\mu_{\mathrm{an}}$. We take a unit vector $\boldsymbol\upsilon$ with the components $\upsilon_k$ and use $|\hat w_{k'-k}|\le\omega_n$ for $k\ne k'$. Separating the diagonal terms and bounding the others by their moduli, we obtain
\begin{equation}
\begin{split}
\langle\boldsymbol\upsilon,E_n\boldsymbol\upsilon\rangle&\le(1-\omega_n)\sum_k\bar E_{kk}|\upsilon_k|^2+\omega_n\sum_{k,k'}\bar E_{kk'}|\upsilon_k||\upsilon_{k'}|\\
&\le(1-\omega_n)\mu_{\mathrm{iso}}+\omega_n(\mu_{\mathrm{iso}}+\mu_{\mathrm{an}}),
\end{split}
\label{eq:mixsplit}
\end{equation}
which is Eq.~\eqref{eq:thmmix}. In the first order
\begin{equation}
E_1=\mathbb E\Lambda=\mu_{\mathrm{iso}}\openone+\mu_{\mathrm{an}}\begin{pmatrix}\mathrm{Re}\,\hat w_1&\mathrm{Im}\,\hat w_1\\\mathrm{Im}\,\hat w_1&-\mathrm{Re}\,\hat w_1\end{pmatrix},
\label{eq:E1mix}
\end{equation}
with the largest eigenvalue $\mu_{\mathrm{iso}}+|\hat w_1|\mu_{\mathrm{an}}$, and the directions of the first order are displacements. For a mixture in which the eigenvalues depend on the phase, $E_n$ is still the average of the efficiency matrices of the settings. Since the largest eigenvalue of an average of matrices is at most the average of their largest eigenvalues, we obtain from Theorem~\ref{thm:single} the bound $\mathbb E\lambda_{\max}(\Lambda)$. $\square$

\emph{Example.} We consider homodyne detection of $\hat x$ and $\hat p$ with equal probabilities near a thermal state. Then $\hat w_m=1$ for even $m$ and $0$ for odd $m$, so $\mathbb E\Lambda=\openone/2$ and every displacement has the efficiency $1/2$. Near the vacuum we have $e_0=e_2=1/2$ in the order $2$ by Eq.~\eqref{eq:homkernel}. The matrix $E_2$ restricted to $\ket0$ and $\ket2$ is then $\tfrac12\big(\begin{smallmatrix}1&1\\1&1\end{smallmatrix}\big)$, with the eigenvalue $1$. Its eigenvector is the squeezing along $x$, which both settings measure with the efficiency $1$.

\subsection{Numerical checks}\label{app:checks}
\label{sec:meth-fisher}The following statements of Sec.~\ref{sec:fisher} were checked with code that does not use the formulas being checked. In this code we integrate $\int\delta^2/f$ for outcome densities calculated directly in the Fock basis (directory \texttt{analyses/fisher} of the repository~\cite{QMIcode}). This code was written and run with an AI coding assistant under our direction, as described in the acknowledgments. We tested Eq.~\eqref{eq:modes} on random directions with Fock support up to photon number $4$ and $\bar n\le5$. For these directions it agrees with the integrals to six digits. The exception is $F_W$ at $\bar n=5$, whose series converges slowly (within $0.3\%$). We have also searched for counterexamples to Eqs.~\eqref{eq:criterion} and~\eqref{eq:thermalbounds} and to the efficiencies at coherent states (Sec.~\ref{sec:fisher-squeezed}) over all directions with Fock support up to photon number $11$, and we found none. For random settings and mixtures of one mode in the orders $n\le8$ we confirmed Theorems~\ref{thm:single} and~\ref{thm:mix} and the bound~\eqref{eq:corhom}. In these tests no efficiency exceeded the bounds by more than $10^{-14}$. We checked Eq.~\eqref{eq:blocks} against direct integrals of the outcome densities of all directions of the orders $n\le4$ near a squeezed thermal state. For the squeezing parameter $0.3$ the two agree to $10^{-10}$. At the larger squeezing parameter $0.6$ the agreement is limited to $10^{-4}$ by the truncation of the Fock basis. For two and three modes we confirmed Eq.~\eqref{eq:Jdiag} to $10^{-12}$ and Theorem~\ref{thm:single} for $240$ pairs of a random state and a random setting in the orders $n\le4$ and $n\le3$, respectively. No efficiency exceeded the bound by more than $10^{-13}$. The spectra of Secs.~\ref{sec:fisher-pure} and~\ref{sec:fisher-tmsv} have been confirmed with symmetric powers of $4\times4$ matrices in the orders $n\le4$ at $r=0.5$ and $r=1$. Here the agreement, limited by the grid of phases, is within $10^{-7}$. The efficiencies of the physical directions, Eq.~\eqref{eq:tmsvloss} and the values for fixed phases have been checked against the Fisher information of the normal outcome distributions, to $10^{-14}$. As a check independent of the Hermite orders, we have also integrated the outcome densities directly in the two-mode Fock basis at $r=0.5$. These integrals agree with Eqs.~\eqref{eq:effpsi} and~\eqref{eq:tmsvloss} to a relative accuracy of $10^{-13}$.

\subsection{Chernoff exponents of a displacement}\label{app:chi2}
Equation~\eqref{eq:chi2} follows from $\delta=f_\rho-f_0$ and $\int(f_\rho-f_0)^2/f_0=\int f_\rho^2/f_0-1$. For the displaced thermal state the heterodyne densities are two normal densities of variance $v+1/2$ per quadrature, and from them we obtain $\xi_Q$. We denote by $\theta_0$ the phase of $\alpha$. For homodyne detection at the phase $\theta$ the overlap $\int\sqrt{f_0f_1}$ of the two outcome densities (the Bhattacharyya coefficient) is then $e^{-2u\cos^2(\theta-\theta_0)}$, and its average over the recorded phase is $e^{-u}I_0(u)$. By symmetry the minimum in Eq.~\eqref{eq:chernoff} is attained at $s=1/2$.

\subsection{Noise channels}\label{app:noise}
A Gaussian channel maps $W$ to the density of $\mathbf Y=T\mathbf X+\mathbf N$ with $\mathbf N\sim\mathcal N(0,N_c)$. The new ratio is then $g'=\mathbb E[g(\mathbf X)|\mathbf Y]$. We apply Eq.~\eqref{eq:condexp} with $H=T$ and $\Xi=V'^{-1}$ and find that the function $e_{\boldsymbol\gamma}$ is mapped to the function $e_{C\boldsymbol\gamma}$ of the normalized quadratures of $V'$, with the matrix $C=V'^{-1/2}TV^{1/2}$. Equation~\eqref{eq:channelmodes} then follows from the Taylor coefficients in Eq.~\eqref{eq:genf}. Under loss $W(\mathbf x)$ is mapped to $\eta^{-1}W(\mathbf x/\sqrt\eta)$ convolved with $\mathcal N(0,\tfrac{1-\eta}{2}\openone)$, and for a thermal state we obtain $C=[\eta v/(\eta\bar n+1/2)]^{1/2}\openone$. For Gaussian noise $C=[v/(v+\sigma^2)]^{1/2}\openone$, the factor of Mehler's formula~\citep{Mehler1866}. Gaussian dephasing with variance $\gamma$ is the average of $\rho\to e^{-i\phi\nhat}\rho e^{i\phi\nhat}$ over $\phi\sim\mathcal N(0,\gamma)$. It leaves a thermal $\rho_0$ unchanged and multiplies the component of angular momentum $l$ by $\mathbb Ee^{il\phi}=e^{-\gamma l^2/2}$.
\end{document}